\documentclass[12pt]{article} 
\pdfoutput=1 
\usepackage[utf8x]{inputenc} 
\usepackage[T1]{fontenc} 
\usepackage{amsmath,amssymb,bbm,bm,commath,mathtools,xfrac} 
\usepackage{xcolor} 
\usepackage{cite} 

\usepackage{titling}
\posttitle{\par\end{center}\bigskip}
\usepackage{authblk}
\usepackage{geometry} 
\usepackage{setspace} 
\usepackage[nottoc,notlof,notlot]{tocbibind} 
\usepackage[titles]{tocloft} 
\usepackage[unicode]{hyperref} 
\hypersetup{bookmarksnumbered=true, bookmarksopen=true, bookmarksopenlevel=2, breaklinks=true, citecolor=blue, colorlinks=true, linkcolor=red, linktoc=page, pdfborder={0 0 0}, pdfstartview=FitH, plainpages=false, unicode=true, urlcolor=blue}

\definecolor{darkgreen}{rgb}{0,0.5,0}

\def\ep{\epsilon}

\DeclareUnicodeCharacter{"0393}{\Gamma}
\DeclareUnicodeCharacter{"0394}{\Delta}
\DeclareUnicodeCharacter{"0398}{\Theta}
\DeclareUnicodeCharacter{"039B}{\Lambda}
\DeclareUnicodeCharacter{"039E}{\Xi}
\DeclareUnicodeCharacter{"03A0}{\Pi}
\DeclareUnicodeCharacter{"03A3}{\Sigma}
\DeclareUnicodeCharacter{"03A5}{\Upsilon}
\DeclareUnicodeCharacter{"03A6}{\Phi}
\DeclareUnicodeCharacter{"03A8}{\Psi}
\DeclareUnicodeCharacter{"03A9}{\Omega}
\DeclareUnicodeCharacter{"03B1}{\alpha}
\DeclareUnicodeCharacter{"03B2}{\beta}
\DeclareUnicodeCharacter{"03B3}{\gamma}
\DeclareUnicodeCharacter{"03B4}{\delta}
\DeclareUnicodeCharacter{"03B5}{\epsilon}
\DeclareUnicodeCharacter{"03B6}{\zeta}
\DeclareUnicodeCharacter{"03B7}{\eta}
\DeclareUnicodeCharacter{"03B8}{\theta}
\DeclareUnicodeCharacter{"03D1}{\vartheta}
\DeclareUnicodeCharacter{"03B9}{\iota}
\DeclareUnicodeCharacter{"03BA}{\kappa}
\DeclareUnicodeCharacter{"03BB}{\lambda}
\DeclareUnicodeCharacter{"03BC}{\mu}
\DeclareUnicodeCharacter{"03BD}{\nu}
\DeclareUnicodeCharacter{"03BE}{\xi}
\DeclareUnicodeCharacter{"03C0}{\pi}
\DeclareUnicodeCharacter{"03C1}{\rho}
\DeclareUnicodeCharacter{"03C3}{\sigma} 
\DeclareUnicodeCharacter{"03C4}{\tau}
\DeclareUnicodeCharacter{"03C5}{\upsilon}
\DeclareUnicodeCharacter{"03C6}{\phi}
\DeclareUnicodeCharacter{"03D5}{\varphi}
\DeclareUnicodeCharacter{"03C7}{\chi}
\DeclareUnicodeCharacter{"03C8}{\psi}
\DeclareUnicodeCharacter{"03C9}{\omega}
\DeclareUnicodeCharacter{"21D0}{\Leftarrow}
\DeclareUnicodeCharacter{"0212B}{\AA}
\DeclareUnicodeCharacter{"00B7}{\cdot}
\DeclareUnicodeCharacter{"00B0}{^{\circ}}
\DeclareUnicodeCharacter{"266A}{\eighthnote}
\DeclareUnicodeCharacter{"266B}{\twonotes}
\PrerenderUnicode{³}\PrerenderUnicode{⁵}
\PrerenderUnicode{×}\PrerenderUnicode{Σ}

\newcommand{\be}{\begin{equation}} \newcommand{\ee}{\end{equation}}
\newcommand{\bea}{\begin{equation} \begin{aligned}} \newcommand{\eea}{\end{aligned} \end{equation}}
\newcommand{\bmu}{\begin{multline}} \newcommand{\emu}{\end{multline}}

\newcommand\eqs[1] {\begin{align}#1\end{align}}
\newcommand\eqsn[1] {\begin{align*}#1\end{align*}}
\newcommand\eqss[1] {\begin{align}\begin{split}#1\end{split}\end{align}}
\newcommand\eqst[1] {\begin{multline}#1\end{multline}}

\newcommand\eqsg[1] {\equ{\begin{aligned}#1\end{aligned}}}
\newcommand\equ[1] {\begin{equation}#1\end{equation}}

\newcommand\half {\tfrac{1}{2}}
\renewcommand\( {\left(}
\renewcommand\) {\right)}

\DeclareMathOperator{\sgn}{sgn}

\newcommand\bC {{\mathbb C}}
\newcommand\bR {{\mathbb R}}

\newcommand\R {{\mathbb R}}

\def\cC{{\mathcal C}}
\def\cI{{\mathcal I}}
\def\cL{{\mathcal L}}
\def\cN{{\mathcal N}}
\def\cQ{{\mathcal Q}}
\def\cR{{\mathcal R}}
\def\cS{{\mathcal S}}
\def\cT{{\mathcal T}}
\def\cW{{\mathcal W}}
\def\cZ{{\mathcal Z}}

\newcommand\W {{\mathcal W}}
\newcommand\Z {{\mathcal Z}}

\newcommand\fg {{\mathfrak g}}
\newcommand\fm {{\mathfrak m}}
\newcommand\fn {{\mathfrak n}}
\newcommand\fq {{\mathfrak q}}

\def\Re{\mathrm{Re}}
\def\Im{\mathrm{Im}}

\newcommand{\tu}{{\tilde{u}}}
\newcommand{\tnu}{{\tilde{\nu}}}
\newcommand{\tgamma}{{\tilde{\gamma}}}

\newcommand\nn {\nonumber\\}

\numberwithin{equation}{section} 
\title{\bf The 5d Superconformal Index at Large $\bm{N}$ \\ and Black Holes}
\author[1]{P. Marcos Crichigno}
\author[2]{Dharmesh Jain}
\affil[1]{Institute for Theoretical Physics, University of Amsterdam, \protect\\
Science Park 904, Postbus 94485, 1090 GL, Amsterdam, The Netherlands \bigskip}
\affil[2]{Department of Theoretical Sciences, \protect\\ S. N. Bose National Centre for Basic Sciences, \protect\\
Block--JD, Sector--III, Salt Lake City, Kolkata 700106, India}
\date{\href{mailto:p.m.crichigno@uva.nl}{\texttt{p.m.crichigno@uva.nl}}\texttt{,~}\href{mailto:dharmesh.jain@bose.res.in}{\texttt{dharmesh.jain@bose.res.in}}}

\begin{document}  

\begin{titlepage}

\maketitle
\pagenumbering{Alph}
\thispagestyle{empty}

\begin{abstract}
We study the large $N$ limit of the superconformal index of a large class of 5d $\cN=1$ superconformal field theories and show it is given by the square of the partition function on the squashed five-sphere.  We show this simple relation implies a Cardy  formula in 5d, which is valid in an ``extended'' regime in which fugacities are finite and  $N$ is large. For theories with weakly coupled gravity duals we conjecture this large $N$ Cardy formula universally accounts for the microscopic entropy of spinning black holes in AdS$_{6}$.  We check this explicitly for known black hole solutions in massive type IIA and type IIB string theory, carrying two angular momenta and one electric charge, and predict the entropy of black holes carrying multiple electric charges, yet to be constructed. We also discuss large $N$ properties of the $S^{3}_{b}\times \Sigma_{\fg}$ partition function, extending previous results to theories with type IIB duals.
\end{abstract}

\end{titlepage}

\pagenumbering{arabic}
\tableofcontents

\section{Introduction and summary}

The large $N$ expansion is one of the most powerful nonperturbative tools in quantum field theory \cite{tHooft:1973alw}. In the  $N\to \infty$ limit,  some theories become exactly solvable or can be recast in terms of a  dual description with a standard perturbative expansion. The large $N$ expansion is, however,  more than a technical tool; it has also led to enormous insights into  the  dynamics of quantum field theories when the number of degrees of freedom at each point is large. A prime example is the  gauge/gravity duality in which a large number of quantum degrees of freedom are reorganized into semiclassical gravitational degrees of freedom.

In this paper, we study universal aspects of 5d $\cN=1$ supersymmetric gauge theories that emerge at large $N$. Our calculations are purely field theoretic and do not assume holography. However, they have interesting  holographic applications, in particular for the counting of microstates of black holes in AdS$_6$.

Given a $d$-dimensional supersymmetric quantum field theory, two important observables are the $S^{d}$ partition function and the superconformal index, or $S^{d-1}\times S^{1}$ partition function with supersymmetric boundary conditions. For the theories we consider, these can be computed exactly via supersymmetric localization, which reduces the path integral to a matrix model.\footnote{See \cite{Pestun:2016zxk} for a review of localization  in various dimensions.}  The physical interpretation of these two observables is quite different. While the (odd-dimensional) sphere partition function is a  measure of the number of degrees of freedom \cite{Jafferis:2011zi,Klebanov:2011gs,Jafferis:2012iv},  the superconformal index counts the number of certain BPS operators in the theory \cite{Kinney:2005ej,Bhattacharya:2008zy}. Correspondingly, one would not expect these two observables, or the  corresponding $N\times N$ matrix models computing them, to be related in any simple way. Although this is true for finite $N$, in the large $N$ limit we will show that for a large class of 5d $\cN=1$ theories the matrix models are in fact closely related.\footnote{As discussed in Section~\ref{sec:The superconformal index} this does not hold for 5d gauge theories with a 6d UV completion. }  As we shall discuss, in a certain region of parameter space where the matrix models are dominated by a single saddle, it follows that the superconformal index is simply the {\it square} of  the sphere partition function:
\equ{\label{SCIS2Intro}
 Z_{S^{4}_{\ep_{1},\ep_{2}}\times S^1_{r}}(\nu_{I}) \approx \left[Z_{S^5_{\vec \omega=(\ep_1,\ep_2, r^{-1})}}(\nu_{I})\right]^2\,.
}
On the LHS the parameters $\nu_{I}$ correspond to chemical potentials for flavor symmetries and $\ep_{1,2}$ are (complexified) squashing parameters of the $S^{4}$. On the RHS the parameters $\nu_{I}$ correspond  to masses for the same flavor symmetries and $\vec \omega$ are squashing parameters of the $S^{5}$. Note the physical meaning of the flavor parameters on both sides is quite different and the map is formal.

As we discuss below, one of the implications of this relation is the universal, large $N$ formula
\equ{\label{5dcardy}
\log Z_{S^{4}_{\ep_{1},\ep_{2}}\times S^1_{r}}(\nu_{I}) \approx -\frac{2 (r\ep_{1}+r\ep_{2}+1)^{3}}{27r^{2}\ep_{1}\ep_{2}}  F_{ S^5}\(\hat m_{I}\)\,,
} 
where $\hat m_{I}=\tfrac{3\nu_{I}}{r\ep_{1}+r\ep_{2}+1}$ and $F_{S^{5}}(\hat{m}_{I})$ is the free energy\footnote{Throughout the paper we define $F=-\log Z$. } of the mass-deformed theory on the round $S^{5}$. This result holds for a vast class of theories engineered in massive type IIA and type IIB string theory and at finite $\ep_{1,2}$ and $r$.  We note the close analogy with the 2d Cardy formula, $\log Z_{S^{1}\times S^{1}_{\beta}}\approx \frac{\pi^{2}}{6\beta}\, c$, where now $F_{ S^5}(\hat m_{I})$ plays the role of the 2d (trial) central charge $c$.

\

As noted, the underlying matrix models controlling the sphere partition function and the superconformal index are quite distinct and we do not expect a simple relation at finite $N$. The universality of the relations above are thus an   emergent property at large $N$.\footnote{See \cite{Benini:2015bwz,Azzurli:2017kxo, Bobev:2017uzs,Crichigno:2018adf, Bobev:2019zmz} for other large $N$ universal relations.} This can be understood holographically as the universality of the Bekenstein-Hawking entropy formula. This formula is universal in the sense that it follows from a weakly coupled, semiclassical, analysis. It therefore holds regardless of how the black hole is realized microscopically, either in massive type IIA or type IIB string theory, or any other setup in a consistent theory of quantum gravity. Now, just like the Bekenstein-Hawking entropy of black holes in AdS$_{3}$ follows universally from the 2d Cardy formula, we expect the entropy of (spinning) black holes in asymptotically AdS$_{6}$ to follow universally from \eqref{5dcardy}.\footnote{As reviewed below, SCFTs in 5d are strongly coupled in the UV and large $N$ is sufficient for a weakly coupled gravity dual.} For black holes that are known one can check this is indeed the case. We conjecture this formula accounts for the entropy of generic spinning, electrically charged, BPS black holes in asymptotically AdS$_{6}$, yet to be constructed, and discuss an example in Section~\ref{sec:Black hole entropy}. It would be interesting to study subleading corrections in the $1/N$ expansion both in the field theory and in the supergravity side.

The formula \eqref{5dcardy} is consistent with the result in \cite{Choi:2019miv}, where the superconformal index of 5d SCFTs engineered in massive type IIA string theory was studied at large $N$ and in a Cardy-like limit, $|\ep_{1,2}|\ll1$.\footnote{See \cite{DiPietro:2014bca,Ardehali:2015hya, Ardehali:2015bla, DiPietro:2016ond,Kim:2019yrz} for 4d Cardy formulas in the large temperature and other Cardy-like limits and \cite{Choi:2018vbz,Choi:2018hmj,ArabiArdehali:2019tdm,Honda:2019cio} for applications to the study of black holes in AdS$_{5}$. } We emphasize the result above holds at finite $\ep_{1,2}$ and $r$. We thus refer to  \eqref{5dcardy} as the ``extended,'' or large $N$,  Cardy formula in 5d. This is reminiscent of the extended Cardy regime for the 2d Cardy formula, in which $N$ is taken to be large but $\beta$ is finite \cite{Hartman:2014oaa}. 

\

We also discuss another  interesting 5d observable: the  $S^3_b\times \Sigma_{\fg}$ partition function \cite{Crichigno:2018adf}, where $\Sigma_{\fg}$ is a topologically twisted Riemann surface of genus $\fg$. Again, we will show that for theories with  UV completions as 5d SCFTs, this partition function is completely controlled, at large $N$, by the mass-deformed partition function on $S^{5}_{\vec \omega}$, with an appropriate map of parameters. This holds for a large class of quiver gauge theories, including theories with massive type IIA duals as well as with type IIB duals. As we discuss in Section~\ref{sec:S3Sigma} our results strongly suggest that the compactification of both classes of 5d SCFTs lead to novel 3d SCFTs which would be interesting to characterize in  detail. 

\

The paper is organized as follows. In Section~\ref{sec:5d} we review basic elements of 5d $\cN=1$ theories and localization results. In Section~\ref{sec:The superconformal index} we study the large $N$ limit of the superconformal index and prove the 5d Cardy formula. In Section~\ref{sec:Black hole entropy} we discuss holographic implications, in particular for the entropy of spinning black holes in AdS$_{6}$. Finally, in Section~\ref{sec:S3Sigma} we discuss the large $N$ limit of the $S^{3}\times \Sigma_{\fg}$ partition function. We provide some technical details  in Appendices~\ref{appAlgebra} and \ref{App:FunIds}.

\section{\texorpdfstring{Review of the 5d Nekrasov partition function and $\bm{S^5}$}{Review of the 5d Nekrasov partition function and S⁵}}
\label{sec:5d}

Five-dimensional gauge theories are trivial in the IR and perturbatively non-renormalizable. At the non-perturbative level, however,  string theory predicts that  various supersymmetric gauge theories are UV completed by nontrivial SCFTs  \cite{Seiberg:1996bd,Intriligator:1997pq}.  Since in 5d the gauge coupling constant squared, $g_{\text{YM}}^2$,  has units of length the UV SCFTs are strongly coupled. There have been important efforts in recent years to classify these 5d SCFTs \cite{Jefferson:2017ahm,Jefferson:2018irk,Bhardwaj:2020gyu} and to develop nonperturbative tools to study them, in particular supersymmetric localization techniques on various backgrounds.\footnote{See Contributions [14-16] in \cite{Pestun:2016zxk} for a review of results in 5d and a complete list of references.}

An important role is played by the instanton partition function \cite{Nekrasov:2002qd} on the $\Omega$-deformed background, $\bC^{2}_{\ep_{1},\ep_{2}}\times S^{1}_{r}$, which serves as a basic building block for various partition functions. For various  manifolds $M_{5}$ the path integral localizes around certain fixed points, where the geometry locally looks like a copy of $\bC^{2}_{\ep_{1},\ep_{2}}\times S^{1}_{r}$. The partition function on $M_{5}$ is then obtained by appropriately gluing  copies of the instanton partition function and (for compact $M_{5}$)  by integrating/summing over gauge configurations: 
\equ{\label{gencopies}
Z_{M_{5}} =\frac{1}{|W_{G}|}\sum_{\fm}\oint_{\cC} \frac{dx}{2\pi i x}\prod_{\ell} \cZ^{\mathit pert}_{\bC^2\times S^1}(x^{(\ell)},y^{(\ell)};\fq_1^{(\ell)};\fq_2^{(\ell)})\cZ^{\mathit inst}_{\bC^2\times S^1}(x^{(\ell)},y^{(\ell)};\fq_1^{(\ell)};\fq_2^{(\ell)})\,,
}
where $x=e^{2\pi i r u}$ are gauge variables,  $y=e^{2\pi i r \nu}$ are flavor fugacities, and $\fq_{1,2}=e^{2\pi i r \ep_{1,2}}$ are the equivariant parameters.  The precise gluing conditions on the parameters, and the sum/integration over gauge variables and the contour $\cC$ depend on the specific choice of $M_{5}$. We will focus on the (squashed) $S^{5}$\cite{Kallen:2012cs,Hosomichi:2012ek,Kallen:2012va,Kim:2012ava,Imamura:2012xg,Imamura:2013xna,Lockhart:2012vp,Kim:2012qf}, $S^{4}\times S^{1}$\cite{Kim:2012gu,Terashima:2012ra,Iqbal:2012xm,Nieri:2013vba}, and $S^{3}\times \Sigma_{\fg}$\cite{Crichigno:2018adf}, with $\Sigma_{\fg}$ a topologically twisted Riemann surface of genus $\fg$.\footnote{Strictly speaking in this approach one can only obtain the case $\fg=0$. An A-model perspective, however, leads to the expression for arbitrary genus \cite{Crichigno:2018adf}. Some partial results were previously obtained in \cite{Fukuda:2012jr,Kawano:2015ssa}. See also \cite{Santilli:2020uht} for recent developments. }  The perturbative part consists of a classical and  1-loop contribution,
\equ{\cZ^{\mathit pert}_{\bC^2\times S^1}=\cZ_{\bC^2\times S^1}^{\mathit classical}\cZ_{\bC^2\times S^1}^{1-loop}\,.
}
As we discuss below, the classical contribution will not play an important role in our setup. The 1-loop term receives contributions from both 5d $\cN=1$ vector and hypermultiplets, which are given by\footnote{Here we follow the regularization and conventions of \cite{Crichigno:2018adf}. The minus sign in front of $x^{±\rho}$  is due to the half-integer shifts of KK momenta by the hypermultiplet R-charge when uplifting the 4d instanton partition function from 4d to 5d \cite{Kim:2012qf}.}
\eqsg{\cZ^{ 1-loop, \mathit{hyp}}_{\bC^2\times S^1}( x,y;\fq_1,\fq_2) &=\prod_{\rho\in \cR}\left[(-x^{\rho}y \, \fq_{1}^{1/2}\fq_{2}^{1/2};\fq_{1},\fq_{2})(-x^{-\rho}y^{-1} \, \fq_{1}^{1/2}\fq_{2}^{1/2};\fq_{1},\fq_{2})\right]^{-1/2}\,, \\
\cZ^{ 1-loop, \mathit{vec}}_{\bC^2\times S^1}( x;\fq_1,\fq_2) &=\prod_{\alpha\in Ad(G)'} \left[(x^{\alpha}\, \fq_{1}\fq_{2};\fq_{1},\fq_{2})(x^{-\alpha};\fq_{1},\fq_{2})\right]^{1/2}\,,
\label{5dblock}}
where $\rho$ runs over weights of the gauge representation $\cR$ and $\alpha$ runs over nonzero roots of the gauge group $G$, we denote $x^{\rho}\equiv e^{2\pi i r \rho(u)}$, and the $(\fq_{1},\fq_{2})$-factorial symbol is defined by
\equ{(z;\fq_{1},\fq_{2})=\prod_{k_{1},k_{2}\geq0}(1-z\fq_{1}^{k_{1}}\fq_{2}^{k_{2}})\,,\qquad |\fq_{1}|<1\, ,|\fq_{2}|<1\,.
}
If either $|\fq_{1}|>1$ or  $|\fq_{2}|>1$ other product representations are used  (see Appendix \ref{App:FunIds}). An important role will be played by the elliptic gamma function, defined as 
\equ{\Gamma(z;\fq_{1},\fq_{2})\equiv\frac{(\fq_{1}\fq_{2}z^{-1};\fq_{1},\fq_{2})}{(z;\fq_{1},\fq_{2})}\,,
\label{DefEllipticGammaT}}
and which satisfies the interesting modular property \cite{FELDER200044}
\equ{Γ\big(z=e^{2\pi i u};\fq_{1}=e^{2\pi iε_1},\fq_{2}=e^{2\pi iε_2}\big)=e^{-i π Q(u;ε_1,ε_2)}\frac{Γ\big(\frac{u}{ε_1};-\frac{1}{ε_1},\frac{ε_2}{ε_1}\big)}{Γ\big(\frac{u-ε_1}{ε_2};-\frac{1}{ε_2},-\frac{ε_1}{ε_2}\big)}\,,
\label{GammaId}}
where $Q$ is a cubic polynomial defined in \eqref{defQA}.

\

The instanton contribution can be explicitly characterized but is more involved. However, as we review below, one expects  sectors with a nonzero instanton number to be subdominant in the large $N$ limit and thus from now on we focus only on the perturbative contribution in the zero instanton sector.

\subsection[\texorpdfstring{Mass deformed squashed $S^{5}$}{Mass deformed squashed S⁵}]{Mass deformed squashed $\bm{S^{5}}$}
\label{sec:Mass deformed squashed S5}

Let us first review how to construct the partition function on the squashed $S^{5}$ (here we follow the review  \cite{Pasquetti:2016dyl}). The squashed $S^5$ is defined by
\equ{\omega_1^2|z_1|^2+\omega_2^2|z_2|^2+\omega_3^2|z_3|^2=1\,,
}
where $z_{1},z_2,z_3$ are coordinates in an embedding $\bC^3$ and  $\vec\omega=(\omega_{1},\omega_{2},\omega_{3})$  are squashing parameters, with units of mass 1. The $S^{5}_{\vec \omega}$ partition function is obtained by gluing three copies, with parameters
\eqss{\label{gluingS5round}
x^{(\ell)}=\begin{cases} e^{2\pi i \,u/\omega_{1}} \\ e^{2\pi i\, u/\omega_{2}} \\ e^{2\pi i \, u/\omega_{3}} \end{cases},\qquad
\fq_{1}^{(\ell)}=\begin{cases} e^{2\pi i  \, \omega_{3}/\omega_{1}} \\ e^{2\pi i  \, \omega_{1}/\omega_{2}} \\ e^{2\pi i  \, \omega_{1}/\omega_{3}} \end{cases},\qquad
\fq_{2}^{(\ell)}=\begin{cases} e^{2\pi i  \, \omega_{2}/\omega_{1}} &\quad \ell=1 \\ e^{2\pi i \, \omega_{3}/\omega_{2}} &\quad \ell=2 \\ e^{2\pi i  \, \omega_{2}/\omega_{3}} &\quad \ell=3 \end{cases}\;,
}
where $u$ is the gauge variable, with units of mass 1. We have set the mass parameters $m$ to zero for now, which can be easily restored later by shifting the gauge variable $u\to u+m$. We consider complexified squashing parameters and assume, for concreteness, 
\equ{\Im\bigg(\frac{{\omega_1}}{{\omega_3}}\bigg)>0\,,\quad\Im\bigg(\frac{{\omega_2}}{{\omega_3}}\bigg)>0\,, \quad \Im\bigg(\frac{{\omega_2}}{{\omega_1}}\bigg)>0\,.
\label{ImOmegasS5}}
Let us consider the contribution from the vector multiplet. With the choice \eqref{ImOmegasS5} and gluing the three copies above one has
\eqst{\cZ^{1-loop,vec}_{S^5_{\vec{\omega}}} =\prod_{\alpha\in Ad(G)'} \frac{\left[(x^{\alpha};\fq_{1}^{-1},\fq_{2}^{-1})(x^{-\alpha}\fq_{1}^{-1}\fq_{2}^{-1};\fq_{1}^{-1},\fq_{2}^{-1})\right]^{1/2}_{(2)}\left[(x^{\alpha}\, \fq_{1}\fq_{2};\fq_{1},\fq_{2})(x^{-\alpha};\fq_{1},\fq_{2})\right]^{1/2}_{(3)}}{\left[(x^{\alpha}\, \fq_{2};\fq_{1}^{-1},\fq_{2})(x^{-\alpha}\,\fq_1^{-1};\fq_{1}^{-1},\fq_{2})\right]^{1/2}_{(1)}} \,,
}
where the label $(\ell)$ indicates that each function is evaluated at the corresponding copy. To extract the leading behavior in the large $N$ limit it is useful to first rewrite this expression by introducing the elliptic gamma function \eqref{DefEllipticGammaT}, as follows:
\equ{\cZ^{1-loop,vec}_{S^5_{\vec{\omega}}}=\prod_{\mathclap{\alpha\in Ad(G)'}} \frac{(x^{-\alpha}\fq_{1}^{-1}\fq_{2}^{-1};\fq_{1}^{-1},\fq_{2}^{-1})_{(2)}(x^{-\alpha};\fq_{1},\fq_{2})_{(3)}}{(x^{-\alpha}\,\fq_1^{-1};\fq_{1}^{-1},\fq_{2})_{(1)}}\left[\frac{\Gamma(x^{\alpha}\, \fq_{2};\fq_{1}^{-1},\fq_{2})_{(1)}}{\Gamma(x^{\alpha};\fq_{1}^{-1},\fq_{2}^{-1})_{(2)}\Gamma(x^{\alpha}\, \fq_{1}\fq_2;\fq_{1},\fq_{2})_{(3)}}\right]^{1/2}·
}
Now, using the modular property \eqref{GammaId} and other shift properties of the elliptic gamma function in Appendix \ref{App:FunIds}, one can show that
\equ{\left[\frac{\Gamma(x^{\alpha}\, \fq_{2};\fq_{1}^{-1},\fq_{2})_{(1)}}{\Gamma(x^{\alpha};\fq_{1}^{-1},\fq_{2}^{-1})_{(2)}\Gamma(x^{\alpha}\, \fq_{1}\fq_2;\fq_{1},\fq_{2})_{(3)}}\right]^{1/2} =e^{\frac{iπ}{6}\(Q_{(2)}+Q_{(3)}-Q_{(1)}\)},
}
where 
\eqst{Q_{(2)}+Q_{(3)}-Q_{(1)} ≡ Q\big(\tfrac{u}{ω_2};-\tfrac{ω_1}{ω_2},-\tfrac{ω_3}{ω_2}\big) +Q\big(\tfrac{u}{ω_3}+\tfrac{ω_1}{ω_3}+\tfrac{ω_2}{ω_3};\tfrac{ω_1}{ω_3},\tfrac{ω_2}{ω_3}\big) -Q\big(\tfrac{u}{ω_1}+\tfrac{ω_2}{ω_1};-\tfrac{ω_3}{ω_1},\tfrac{ω_2}{ω_1}\big) \\
=\frac{u^3}{\omega_1\omega_2\omega_3} +\frac{3\omega_{\text{tot}}}{2 \omega_1\omega_2\omega_3}u^2 +\frac{\omega_{\text{tot}}^2 +\omega_1\omega_2+\omega_1\omega_3+\omega_2\omega_3 }{2\omega_1\omega_2\omega_3}u + \frac{\omega_{\text{tot}}(\omega_1\omega_2+\omega_1\omega_3+\omega_2\omega_3)}{4\omega_1\omega_2\omega_3}\,,
\label{threeQs}}
with $\omega_{\text{tot}}\equiv \omega_{1}+\omega_{2}+\omega_{3}$ and we used the expression for $Q$ in   \eqref{defQA}. It is easy to check that this combination is  precisely a  Bernoulli polynomial, $B_{3,3}\,$, given in \eqref{defB}. Thus, the 1-loop contribution from the vector can be written as 
\equ{\cZ^{1-loop,vec}_{S^5_{\vec{\omega}}} =\prod_{\alpha\in Ad(G)'} e^{-\frac{iπ}{6}B_{3,3}(-\alpha(u)|\vec{\omega})}∏_{\ell=1}^3(x^{-\alpha};\fq_1,\fq_2)_{(\ell)}\,.
\label{vecS5lN}}
The contribution from the hypermultiplet can be obtained from that of the vector, by the replacement  $α(u)→ρ(u)+m-\frac{\omega_{\text{tot}}}{2}$. Then, the total 1-loop contribution can be written as
\equ{\cZ^{1-loop}_{S^5_{\vec{\omega}}} =e^{\Psi(u,m|\vec \omega)}\prod_{\alpha\in Ad(G)'}\prod_{ρ\in\cR}\, ∏_{\ell=1}^3\frac{(x^{-\alpha};\fq_1,\fq_2)_{(\ell)}}{\big(-x^{-ρ}y^{-1}\fq_1^{1/2}\fq_2^{1/2};\fq_1,\fq_2\big)_{(\ell)}}\,,
\label{1-loopS5}}
where 
\equ{\label{defPsi}
e^{\Psi(u,m|\vec \omega)}\equiv \prod_{\alpha\in Ad(G)'}\prod_{ρ\in\cR}e^{-\frac{iπ}{6}B_{3,3}(-\alpha(u)|\vec{\omega})+\frac{iπ}{6}B_{3,3}\(-ρ(u)-m+\frac{\omega_{\text{tot}}}{2}|\vec{\omega}\)}\,.
}

Before studying the large $N$ limit of \eqref{1-loopS5}, which will be dominated by the prefactor $e^{\Psi}$, we review the role of the classical and instanton contributions at large $N$. This is  subtle and depends on the theory under consideration. On the one hand, to access the UV fixed point one would want to take the strong coupling limit, $ g_{\text{YM}}^{2}\gg R$, where $R$ is the size of the $S^{5}$. In this limit the classical Yang-Mills action vanishes and gives no contribution to the partition function.  On the other hand, instanton effects generally become important at strong coupling, and should therefore be taken into account in describing the UV physics. However, the instanton sum is controlled not by the bare gauge coupling constant but by the  effective gauge coupling constant, which is shifted by 1-loop effects and varies over the Coulomb branch \cite{Seiberg:1996bd}. Taking this into account, the upshot of the analysis in \cite{Jafferis:2012iv} is that for a large class of theories, including those arising from D4-D8-O8 systems in massive type IIA string theory \cite{Seiberg:1996bd} and those arising from $(p,q)$-fivebrane webs in type IIB string theory \cite{Aharony:1997ju,Kol:1997fv,Aharony:1997bh}, there is a large $N$ regime where both the classical and instanton contributions are suppressed:
\equ{\cZ^{\mathit classical}_{S^5_{\vec \omega}} \approx 1\,, \qquad \cZ^{\mathit inst}_{S^5_{\vec \omega}} \approx 1\,.
}
It is important to note that this is not  the case for all 5d gauge theories, in particular the maximal SYM theory. For this theory, in the regime in which instantons are suppressed, the classical contribution is of the same order in $N$ as the 1-loop contribution and thus cannot be ignored \cite{Kallen:2012zn}. In fact, in this case  $g_{\text{YM}}^2=2\pi \beta$ is identified with an emergent circle of radius $\beta$ and the gauge theory is UV completed by the 6d $(2,0)$ theory on $S^{1}_{\beta}$ rather than by a 5d SCFT \cite{Douglas:2010iu,Lambert:2010iw}.\footnote{This is also the case for other 5d $\cN=1$ gauge theories such as the Seiberg theory with $N_{f}=8$ or theories of class $\mathcal S_{k}$, which have UV completions as 6d theories. All such theories are excluded in our analysis.  } 

In what follows we restrict ourselves to theories for which the classical as well as instanton contributions in the UV are suppressed at large $N$. Then, the large $N$ limit of the partition function is dominated by the large $N$ behavior of the 1-loop determinants \eqref{1-loopS5}. Studies of the $S^5$ partition function reveal  that at large $N$ the matrix model is dominated by complex saddles, $u_i^\ast$, which are large and imaginary.\footnote{See \cite{Jafferis:2012iv} for theories in massive type IIA and  \cite{DHoker:2016ysh,Uhlemann:2019ypp} for theories in type IIB.}  Let us write $u_i= i \sigma_i$ with $\sigma_i\in \bR$ and take the eigenvalues to be ordered as $\sigma_{1}>\sigma_{2}>\ldots>\sigma_{N}$.  Consider then, say, the contributions from the positive roots, $α>0$, and in  the limit $\sigma_i→-∞$.  Assuming $\Re\, \omega_{i}>0$ the $(\fq_{1},\fq_{2})$-Pochhammer  symbols simply become 1 and the leading contribution arises from the exponential factor in \eqref{1-loopS5}. For negative roots, or in the limit $\sigma_i→∞$, one  performs the analogous manipulations to bring the expression into the same form as \eqref{1-loopS5} but with $\sigma_i \to -\sigma_i$ and the large $N$ limit is again controlled by the prefactor, $e^{\Psi(-i \sigma,-m|\vec \omega)}$. 

For a general theory with gauge group $G$ and hypermultiplets in gauge representations $\cR_{I}$, the final result is that in the large $N$ limit,  
\equ{\label{integrandS5largeN}
\cZ_{S^5_{\vec{\omega}}} \approx \prod_{\alpha\in Ad(G)'}\prod_{I}\prod_{ρ\in\cR_{I}}\,e^{-F_{V}\(\alpha(\sigma)|\vec \omega\)-F_{H}\(\rho(\sigma)-im_{I}|\vec \omega\)} \,,
}
where $m_{I}$ are mass parameters for each flavor symmetry acting on the hypermultiplet and 
\eqss{F_{V}(x|\vec \omega)= \,& \sgn(x)\(\frac{\pi}{6 \omega_{1}\omega_{2}\omega_{3}}x^{3}-\frac{\pi(\omega_{\text{tot}}^{2}+\omega_{1}\omega_{2}+\omega_{1}\omega_{3}+\omega_{2}\omega_{3} )}{12 \omega_{1}\omega_{2}\omega_{3}}x\),\\
F_{H}(x|\vec \omega)= \,&  -\sgn(x)\(\frac{\pi}{6 \omega_{1}\omega_{2}\omega_{3}}x^{3}+\frac{\pi(\omega_{1}^{2}+\omega_{2}^{2}+\omega_{3}^{2} )}{24 \omega_{1}\omega_{2}\omega_{3}}x\).
\label{fVHS5}}
This result was obtained in \cite{Imamura:2013xna,Alday:2014bta} by similar manipulations. 

Let us briefly set $m_{I}=0$. Then, it was shown that for a large class of theories at large $N$ the dependence on the squashing parameters factors out in a simple way \cite{Alday:2014rxa,Alday:2014bta}:
\equ{F_{S^5_{\vec{\omega}}}=\frac{\omega_{\text{tot}}^3}{27\omega_1\omega_2\omega_3}F_{S^5}\,,
}
where $F_{S^5}$ is the free energy of the theory on the round $S^{5}$ of unit radius, $\vec \omega=(1,1,1)$.  Let us see exactly for which class of theories this factorization holds.  Rescaling the gauge variable $x→\frac{\omega_{\text{tot}}}{3}x$ we have
\eqsg{F_{V}(x|\vec{\omega}) &=\frac{\omega_{\text{tot}}^3}{27\omega_1\omega_2\omega_3}\sgn(x)\left[\frac{π}{6}x^3 -\frac{3π}{4}\(1 +f(\vec \omega)\)x\right],\\
F_{H}(x|\vec{\omega}) &=\frac{\omega_{\text{tot}}^3}{27\omega_1\omega_2\omega_3}\sgn(x)\left[-\frac{π}{6}x^3 -\frac{3π}{4}\(\tfrac{1}{2} -f(\vec \omega)\)x\right],
}
where $f(\vec \omega)\equiv(\omega_1\omega_2+\omega_1\omega_3+\omega_2\omega_3)\omega_{\text{tot}}^{-2}$.
Thus, the condition for the overall factorization of the squashing parameters is that when the contributions from the vector and hypermultiplet are combined,  the term proportional to $f(\vec \omega)$ is subleading in the $1/N$ expansion, i.e., 
\eqss{\label{constquiver}
\lim_{N\to \infty}\frac{\sum_{\alpha\in Ad(G)'}|\alpha(\sigma)|-\sum_I\sum_{\rho\in R_I}|\rho(\sigma)|}{\sum_{\alpha\in Ad(G)'}|\alpha(\sigma)|^3-\sum_I\sum_{\rho\in R_I}|\rho(\sigma)|^3} = 0\,,\\
\lim_{N\to \infty}\frac{\sum_{\alpha\in Ad(G)'}|\alpha(\sigma)|-\sum_I\sum_{\rho\in R_I}|\rho(\sigma)|}{\sum_{\alpha\in Ad(G)'}|\alpha(\sigma)|+\frac12\sum_I\sum_{\rho\in R_I}|\rho(\sigma)|} = 0\,.
}
Requiring that this holds everywhere on the Coulomb branch leads to a constraint on the quiver, which depends on the type of gauge groups and matter representations. This condition is satisfied for theories constructed both in massive type IIA and type IIB string theory and we shall assume it holds for all theories we consider.\footnote{See \cite{Crichigno:2018adf} for explicit examples with various gauge groups. } As we shall discuss, this condition also appears in the large $N$ analysis of the superconformal index and the $S^{3}_b\times \Sigma_{\fg}$ partition function.

Turning mass parameters $m_{I}$ back on, we see that for any such theories at large $N$ the squashed $S^{5}$ matrix model is dominated by the saddle configuration,
\equ{\label{saddleS5}
\{\hat \sigma_{i}^{\ast}\}=\left\{\hat\sigma_{i}\, \Big|\,\frac{\partial F(\hat\sigma)}{\partial \hat\sigma_{i}}=0\right\},
}
where
\equ{F(\hat\sigma)=\frac{\omega_{\text{tot}}^3}{27\omega_1\omega_2\omega_3}\Bigg[\sum_{\alpha\in Ad(G)'}F_{V}\(\alpha(\hat\sigma)\)+\sum_{I}\sum_{\rho\in R_{I}}F_{H}\(\rho(\hat\sigma)- i \hat m_{I}\)\Bigg]\,,
}
with $\hat\sigma \equiv \frac{3}{\omega_{\text{tot}}}\sigma$, $\hat m_{I}  \equiv \frac{3}{\omega_{\text{tot}}} m_{I}$, and  $F_{V,H}(x)\equiv F_{V,H}(x|1,1,1)$, which control the round $S^{5}$ partition function \cite{Jafferis:2012iv}. One may solve these saddle equations explicitly for specific theories but we will not need the explicit solutions here.\footnote{See \cite{Jafferis:2012iv} for SCFTs in massive type IIA,  \cite{Uhlemann:2019ypp} for theories in IIB, and \cite{Chang:2017mxc,Chang:2017cdx,Gutperle:2018axv,Hosseini:2019and} for the inclusion of mass deformations.}  Thus, we have shown the relation between the mass-deformed partition functions
\equ{\label{roundSquashed}
F_{S^5_{\vec{\omega}}}\(m_{I}\)=\frac{\omega_{\text{tot}}^3}{27\omega_1\omega_2\omega_3}F_{S^5}\(\tfrac{3m_{I}}{\omega_{\text{tot}}} \)\,.
}
This relation will be useful in the next sections.

\section{\texorpdfstring{The superconformal index at large $\bm{N}$}{The superconformal index at large N}}
\label{sec:The superconformal index}

We now turn to the main subject of the paper, the large $N$ limit of the superconformal index. The superconformal index is defined as \cite{Bhattacharya:2008zy,Kim:2012gu}
\equ{\label{Hindex}
\cI_{S^{4}}(y_{I},q;\fq_{1},\fq_{2})=\text{Tr} \, (-1)^F e^{-\beta \{\cQ,\cQ^{\dagger}\}}\, \fq_{1}^{J_{1}+R}\, \fq_{2}^{J_{2}+R} \, y_I^{Q_{I}} \,q^k\,,
}
where the trace is taken over the Hilbert space of the theory quantized on $S^{4}$ and $F$ is the fermion number.\footnote{Our specific choice of fermion number is $(-1)^{F}=e^{2\pi i R}$, as in \cite{Cordova:2016uwk} for 4d $\cN=2$ theories. More conventional choices such as $(-1)^{F}=e^{\pi i (J_{1}+J_{2})}$ are related to this by (complex) redefinitions of chemical potentials. }  The operators $J_{1,2}$ are the generators of the $SO(2)\times SO(2)\subset SO(5)$ Cartan of  rotations of the $S^{4}$, $R$ is the generator of the Cartan of the $SU(2)_{R}$ R-symmetry, and $Q_{I}$ are the generators in the Cartan of the flavor group, all with their corresponding fugacities. The charges $R$ and $J_{1,2}$ are quantized to be half-integers.  The parameter $q$ is a fugacity for the topological $U(1)$ symmetry of 5d theories, with $k$ the corresponding instanton number, which will play no role in the large $N$ limit.  By standard arguments, the index receives contributions only from $\tfrac18$-BPS states,  annihilated by the supercharges $\cQ$ and $\cQ^{\dagger}=\cS$ and is thus independent of the parameter $\beta$. The anticommutator reads $\{\cQ,\cQ^{\dagger}\}=\Delta -J_{1}-J_{2}-3R$, with $\Delta$ the conformal dimension. It additionally follows from the 5d superconformal algebra (see Appendix~\ref{appAlgebra}) that the states contributing to the index satisfy 
\equ{\label{poscharges}
J_1+3 R\geq 0\,, \qquad J_2+3 R\geq 0\,, \qquad J_1+J_2\geq 0\,.
}
States saturating any of these inequalities preserve additional supersymmetry. 

\

The index can alternatively be expressed as a path integral of the Euclidean action on $S^4_{ε_1,ε_2}\times S^1_{r}$, where $\ep_{1,2}$ are identified with complexified squashing parameters of the $S^{4}$ and $r$ is the radius of the $S^{1}$. The path integral on this background can be evaluated by supersymmetric localization, which localizes it on the north and south pole of the $S^4_{\ep_{1},\ep_{2}}$, where the space locally looks like the 5d $\Omega$-deformed background with parameters $\pm \ep_{1,2}$. The superconformal index thus takes the form \eqref{gencopies} with two copies of  Nekrasov partition functions with parameters (see \cite{Pasquetti:2016dyl} for a review):
\eqss{\label{gluingSCI}
x^{(\ell)}=\begin{cases} e^{2\pi i ru} \\ e^{-2\pi i ru} \end{cases}, \qquad 
y^{(\ell)}=\begin{cases} e^{2\pi i r \nu} \\ e^{-2\pi i r \nu} \end{cases},\qquad 
 \fq_{i}^{(\ell)}=\begin{cases} \fq_{i} & \qquad \ell=1 \\ \fq_{i}^{-1} &\qquad \ell=2 \end{cases}\;,
}
where $\fq_{i}\equiv e^{2\pi i r \epsilon_{i}}$. We will work with complexified fugacities, with the identifications
\equ{u\sim u+\frac{1}{r}\,,\qquad \nu\sim \nu+\frac{1}{r}\,,\qquad \ep_{1,2}\sim \ep_{1,2}+\frac{1}{r}\,·
}
We then restrict the real part of fugacities to the domains $0\leq \Re \, u <\frac1r$, $0\leq \Re \, \nu <\frac1r$, and $0\leq \Re\, \ep_{1,2}<\frac1r\,·$ Gluing these two copies the classical contribution cancels out and the perturbative contribution is given entirely by the 1-loop determinants. We shall assume that instanton contributions are suppressed at large $N$ and thus the dominant contribution is entirely from the perturbative sector.\footnote{The instanton contribution at large $N$ was studied in detail in \cite{Choi:2019miv} for the Seiberg theories, showing they are indeed suppressed at large $N$, as for the sphere partition function. We assume this holds  more generally, which is supported by our results. }
 We take
\equ{\label{choiceepsilons}
|\fq_{1}|<1\,,\quad |\fq_{2}|<1\qquad \Leftrightarrow \qquad \Im \, ε_1 >0\,,\quad \Im \, ε_2>0\,.
}
Then, from \eqref{5dblock} the vector contribution  reads
\eqs{\Z^{1-loop,vec}_{S^4_{\ep_1,\ep_2}×S^1_r} &=\prod_{\alpha\in Ad(G)'}\left[(x^{\alpha}\, \fq_{1}\fq_{2};\fq_{1},\fq_{2})(x^{-\alpha};\fq_{1},\fq_{2})\right]^{1/2}_{(1)}\left[(x^{\alpha};\fq_{1}^{-1},\fq_{2}^{-1})(x^{-\alpha}\fq_{1}^{-1}\fq_{2}^{-1};\fq_{1}^{-1},\fq_{2}^{-1})\right]^{1/2}_{(2)} \nn
&=\prod_{\alpha\in Ad(G)'}\frac{(x^{-\alpha};\fq_{1},\fq_{2})^2}{Γ(x^{\alpha}\, \fq_{1}\fq_{2};\fq_{1},\fq_{2})}\,·}
The contribution from the hypermultiplet is obtained by replacing $r\alpha(u)\to r\rho(u)+r\nu-\tfrac12(r\ep_1+r\ep_2+1)$ and taking the inverse. The total perturbative partition function is then
\eqs{Z^{\mathit pert}_{S^4_{\ep_1,\ep_2}×S^1_r} =\frac{1}{|W_G|}\oint du_i \, e^{\hat \Psi(u,\nu|\vec \omega)}\, \prod_{\alpha\in Ad(G)'}  \prod_{\rho\in \cR}  \frac{(x^{-\alpha};\fq_{1},\fq_{2})^2}{(-x^{-\rho}y^{-1}\fq_{1}^{1/2}\fq_2^{1/2};\fq_{1},\fq_{2})^2}\,,
\label{1-loopTotP}}
where
\equ{\label{psiSCI}
e^{\hat \Psi(u,\nu|\vec \omega)}\equiv  \prod_{\alpha\in Ad(G)'}  \prod_{\rho\in \cR}\frac{Γ(-x^{\rho}y\, \fq_{1}^{1/2}\fq_{2}^{1/2};\fq_{1},\fq_{2})}{Γ(x^{\alpha}\, \fq_{1}\fq_{2};\fq_{1},\fq_{2})}\,,
}
and the integration contour is around the unit circles, $|e^{2\pi i r u_{i}}|=1$, for each $i$. To evaluate this integral at large $N$ we  perform a saddle-point approximation. 

Now, to analyze the large $N$ limit we write $u_i = i \sigma_i$ and expand the functions for  $\sigma_i\to \pm \infty$, as in the case of $S^{5}$ in Section~\ref{sec:Mass deformed squashed S5}. Arguing as for the case of $S^5$, the $(\fq_1,\fq_2)$-Pochhammer symbols in \eqref{1-loopTotP} are subleading and the 1-loop determinants are dominated by the prefactor  \eqref{psiSCI}, which in this case is more complicated than  its $S^5$ counterpart \eqref{defPsi}. We derive the asymptotics of the elliptic gamma function  in Appendix~\ref{sec:AsymptoticsEG} and which turns out to be dominated by the Bernoulli polynomial $B_{3,3}$. The basic idea is to relate this function to Barnes' triple gamma function via \eqref{triplegammaA}, whose asymptotics  was studied in \cite{RUIJSENAARS2000107} and is reproduced in \eqref{expPsi3}. Considering the positive roots and weights we can write the function in \eqref{psiSCI} in the $u\to -i \infty$ limit as
\equ{\frac{Γ(-x^{\rho}y\, \fq_{1}^{1/2}\fq_{2}^{1/2};\fq_{1},\fq_{2})}{Γ(x^{\alpha}\, \fq_{1}\fq_{2};\fq_{1},\fq_{2})} ≈ e^{-\frac{iπ}{3}B_{3,3}(-\alpha(u)|\,ε_1,ε_2,r^{-1})+\frac{iπ}{3}B_{3,3}(-\rho(u)-\nu-\frac{1}{2}(\ep_1+\ep_2+r^{-1})|\,ε_1,ε_2,r^{-1})}\,,
\label{gammavec}}
where we have assumed the quiver constraint \eqref{constquiver}.  For negative roots and weights, or in the limit $u\to i \infty$, one simply repeats the manipulations above with $u\to -u$. For a general theory with gauge group $G$ and hypermultiplets in gauge representations $\mathcal{R}_{I}$ the superconformal index at large $N$ is then given by
\equ{\label{integrandSCIlargeN}
\cZ_{S^4_{\ep_1,\ep_2}×S^1_r} \approx \prod_{\alpha\in Ad(G)'}\prod_{I}\prod_{ρ\in\cR_{I}}\,e^{-2F_{V}\(\alpha(\sigma)|\, ε_1,ε_2,r^{-1}\)-2F_{H}\(\rho(\sigma)-i\nu_{I}|\, ε_1,ε_2,r^{-1}\)} \,,
}
where the functions $F_{V}$ and $F_{H}$ were defined in \eqref{fVHS5} and the $\nu_{I}$ are fugacities for the flavor symmetries acting on each hypermultiplet.  Note this is precisely the square of the integrand  \eqref{integrandS5largeN} for  the squashed sphere at large $N$, with the identifications
\equ{\label{idomega}
\vec \omega = (\ep_{1},\ep_{2},r^{-1})\,,\qquad m_{I}=\nu_{I}\,.
}
The fugacity parameters, $\nu_{I}$, are mapped to masses, $m_{I}$, for the same symmetries in the sphere partition function. These have different physical interpretations on both sides and thus the map is formal. We also note that the mapping of squashing parameters is compatible with the assumptions \eqref{ImOmegasS5} and \eqref{choiceepsilons}.\footnote{One can repeat the analysis for all other choices and the final conclusion is unchanged. }  Then, we have shown that, to leading order in $N$, the matrix models for the superconformal index and sphere partition function are related by
\eqs{\cZ_{S^4_{ε_1,ε_2}×S^1_r}(i\sigma,\nu_{I})\approx  \left[\cZ_{S^5_{\vec \omega=(ε_1,ε_2,r^{-1})}}(i\sigma,\nu_{I})\right]^2\,.
\label{SCIS5sq}
}
Given this relation, it is clear that the saddle \eqref{saddleS5} for the sphere partition function in this limit is also a saddle for the superconformal index. Furthermore, when there is a  single saddle dominating the matrix models,  we arrive at the rather remarkable conclusion that the superconformal index is the square of  the (mass-deformed) sphere partition function:
\equ{{Z_{S^4_{ε_1,ε_2}×S^1_r}(\nu_{I}) \approx \left[ Z_{S^5_{\vec \omega=(ε_1,ε_2,r^{-1})}}(\nu_{I})\right]^2}\,,
\label{square}}
to leading order in the $1/N$ expansion.

\subsection[\texorpdfstring{A large $N$ Cardy formula}{A large N Cardy formula}]{A large $\bm{N}$ Cardy formula}
\label{sec:A large N Cardy formula}

We now point out an interesting consequence of  \eqref{square}. As reviewed in Section~\ref{sec:Mass deformed squashed S5}, for any quiver gauge theory satisfying \eqref{constquiver}, the squashed and round sphere partition functions are related at large $N$ by \eqref{roundSquashed}.   For any such theory, it then follows that
\equ{\label{5dcardyT}
\log Z_{S^{4}_{\ep_{1},\ep_{2}}\times S^1_{r}}(\nu_{I}) \approx -\frac{2 (r\ep_{1}+r\ep_{2}+1)^{3}}{27r^{2}\ep_{1}\ep_{2}}  F_{ S^5}\(\hat m_{I}\)\,,
} 
where $\hat m_{I}=\tfrac{3\nu_{I}}{r\ep_{1}+r\ep_{2}+1}\,·$   This  result  encapsulates and extends a number of observations in the literature.  For the case of theories with massive IIA duals the formula \eqref{5dcardyT} is consistent with the expression for the superconformal index derived in \cite{Choi:2019miv}, which was studied in the large $N$ and Cardy-like limit $|\ep_{1,2}|\ll1$. 

We emphasize that our result holds for a vast class of theories, including those arising in massive type IIA as well as type IIB string theory, and  for finite $\ep_{1,2}$ and $r$.\footnote{This was anticipated from holography in \cite{Bobev:2019zmz}, based on the results in \cite{Choi:2018fdc,Zaffaroni:2019dhb,Cassani:2019mms}.}  As discussed in the Introduction, this is reminiscent of the {\it extended} regime of the 2d Cardy formula \cite{Hartman:2014oaa}. We conjecture that the 5d extended Cardy formula universally accounts for the entropy of spinning black holes in AdS$_{6}$, just like the 2d Cardy formula universally accounts for the entropy of black holes in AdS$_{3}$. It would be interesting to study whether there is an underlying ``modular property'' which explains the 5d Cardy formula, perhaps along similar lines to what has been observed for 4d $\cN=4$ SYM in \cite{Benini:2018ywd,Cabo-Bizet:2019eaf} and more recently for $\cN=1$ theories \cite{Gadde:2020bov}. This may also be useful in establishing the precise regime of validity of \eqref{5dcardyT} away from the Cardy-like limit. 

\

We comment in passing that a relation analogous to \eqref{square} has been shown to hold in 3d, relating the large $N$ superconformal index, $Z_{S^{2}_{\omega}\times S^{1}}$, in the Cardy-like limit $|\omega|\ll1$, to the  squashed $S^3$ partition function in the interesting paper \cite{Choi:2019dfu}. The results discussed here, and the holographic arguments in \cite{Bobev:2019zmz}, suggest that at large $N$ the 3d relation may hold in an extended regime as well. It would be interesting to study this in detail. 

\paragraph{Comment on other possible large $\bm{N}$ contributions.} One should keep in mind that the analysis carried out above does not necessarily reveal {\it all} complex saddles contributing to the superconformal index at large $N$. For instance, a similar matrix model -- in terms of elliptic gamma functions -- controls the supersymmetric index in 4d. An alternative representation of the index in terms of a Bethe ansatz formula \cite{Benini:2018ywd}, however,  reveals the existence of additional eigenvalue configurations at large $N$ which are not directly visible as saddles in the original representation of the index but are nonetheless important. This has also been understood in a different approach in \cite{Cabo-Bizet:2019eaf}. Thus, it is possible that a similar reformulation of the 5d supersymmetric index may reveal such additional eigenvalue configurations.  These other configurations, if they exist, may in fact become dominant in certain regions of the space of complex fugacities. This would place a constraint on the region in fugacity space in which \eqref{5dcardyT} holds.  A more systematic approach to  studying additional complex saddles could be carried out, for instance, by Picard-Lefschetz theory. It would be interesting to study if any such methods reveal additional configurations and their precise region of dominance,  which  lies beyond the scope of this paper.  Nonetheless, the saddle \eqref{saddleS5}, which leads to the contribution \eqref{5dcardyT} in the index, always exists even if it may not be the dominant contribution in all of parameter space. As we discuss in Section~\ref{sec:Black hole entropy} this saddle has a holographic interpretation as a single-center black hole.

\subsection{Microcanonical partition function}
\label{sec:microcanonicalpf}

Let us briefly review how to extract the microscopic entropy. From the Hamiltonian representation of the index \eqref{Hindex} and denoting $\fq_{1,2}=e^{2\pi i r \ep_{1,2}}\equiv e^{\omega_{1,2}}$, with $\Re\, \omega_{1,2} <0$, it follows that it takes the form
\equ{\label{SCIdeg}
Z_{S^4_{\omega_1,\omega_2}\times S^1}(\nu_{I})=\sum_{\text{BPS states}}\Omega(J_1,J_2,Q_{R},Q_I)\, e^{\omega_1 J_1} e^{\omega_2 J_2} e^{\nu_{R}Q_{R}}e^{\nu_I Q_I}\,,
} 
where we defined  the R-symmetry fugacity $\nu_{R}=\omega_{1}+\omega_{2}+ 2\pi i n$, with $n=\pm 1$, and $\Omega(J_1,J_2,Q_{R},Q_I)$ is the degeneracy of  $\frac{1}{8}$-BPS states with angular momenta $J_{1,2}$, R-charge $Q_{R}$, and flavor charges $Q_I$.  In the canonical ensemble the fugacities are fixed and all $\tfrac18$-BPS states with charges satisfying \eqref{poscharges} contribute to the partition function. 

Since we are interested in the degeneracy of a particular state with given charges $(J_{1,2},Q_R,Q_I)$, we go to the microcanonical ensemble. This is obtained by
\equ{\Omega(J_1,J_2,Q_{R},Q_I)= \int_{\cC} \frac{d\nu_I}{2\pi i} \frac{d\nu_R}{2\pi i}   \frac{d\omega_{1,2}}{2\pi i}  \frac{d\lambda}{2\pi i} \, Z_{S^4_{\omega_1,\omega_2}\times S^1} (\nu_{I})\, e^{-\omega_1 J_1} e^{-\omega_2 J_2} e^{
-\nu_{R}Q_{R}}e^{-\nu_I Q_I}e^{\lambda(\omega_{1}+\omega_{2}-\nu_R+ 2\pi i n)}\,,
}
where  we temporarily relaxed the supersymmetry constraint, imposing it by the Lagrange multiplier $\lambda$. The entropy is  defined, as usual, by the logarithm of the microcanonical partition function. The contour $\cC$ is chosen so that the integral is convergent and we assume it can be deformed to pass through saddles of the integrand. Then, in the large $N$ limit the integral can be evaluated by the saddle-point approximation,
\equ{\label{FTentropy}
S\equiv  \log \Omega \approx \log Z_{S^4_{\omega_1,\omega_2}\times S^1}(\nu_{I})   - \omega_1 J_1-\omega_2 J_2 -\nu_{R} Q_{R}-\nu_{I}Q_{I}+\lambda(\omega_{1}+\omega_{2}-\nu_{R} + 2\pi i n) \,,
}
subject to the usual Legendre transform relations. We can now consider the contribution from the saddle leading to \eqref{5dcardyT}
\equ{\label{FTentropyE}
S\approx  - \frac{\nu_{R}^{3}}{27i \pi \omega_{1}\omega_{2}}  F_{ S^5}(\nu_{I})  - \omega_1 J_1-\omega_2 J_2 -\nu_{R} Q_{R}-\nu_{I}Q_{I}+\lambda(\omega_{1}+\omega_{2}-\nu_{R} + 2\pi i n)\,.
}

\paragraph{Universal sector.} Setting all flavor fugacities to zero, $\nu_{I}=0$, is special. In this case, the superconformal index is concerned only with the degeneracy of BPS states with fixed R-charge $Q_R$ and angular momenta $J_{1,2}$ (and hence fixed $\Delta$) and is oblivious to any flavor quantum numbers the states may carry. Since the stress-energy tensor multiplet is a universal sector of all SCFTs, and following the terminology of \cite{Bobev:2019zmz,Bobev:2017uzs}, we refer to this as the universal case. Using \eqref{5dcardyT} the saddle equations read
\eqsn{0=\,&\lambda-J_1 +\frac{1}{i\pi} \frac{\varphi^{3}}{\omega_{1}^2\omega_{2}} F_{S^{5}}  \,,&& 0= \lambda-J_2 +\frac{1}{i\pi} \frac{\varphi^{3}}{\omega_{1}\omega_{2}^2} F_{S^{5}}  \,,\\
0=\,&  \lambda+Q_R +\frac{1}{i\pi} \frac{\varphi^{2}}{\omega_{1}\omega_{2}} F_{S^{5}} \,, && 0=\omega_{1}+\omega_{2}-3\varphi + 2\pi i n\,,
}
where $F_{S^{5}}$ is the free energy of the theory on the round $S^{5}$ with all mass-deformation parameters set to zero and we defined $\nu_{R}=3\varphi$.  Then, at the saddle
\equ{S= 2\pi i n \lambda\,,
}
and the entropy is determined by the Lagrange multiplier $\lambda$.  To determine $\lambda$ we take products of the equations above,
\equ{(\lambda-J_1)(\lambda-J_2) = -\frac{i \pi }{F_{S^5}}(\lambda+Q_R)^3\,.
}
The solution to this cubic equation for $\lambda$ is generically complex and thus the entropy is not real. Demanding a purely imaginary $\lambda$ and assuming $J_{1,2}$ and $Q_R$ to be real, the complex conjugate equation reads
\equ{(\lambda+J_1)(\lambda+J_2) = -\frac{i \pi }{F_{S^5}}(\lambda-Q_R)^3\,.
}
Taking the difference of the two equations  gives a quadratic equation for $\lambda$, with two roots. This gives 
\equ{S= 2\pi n F_{S^{5}}\frac{J_1+J_2\pm\sqrt{(J_1+J_2)^2+\frac{12\pi^2Q_R^4}{F_{S^5}^2}}}{3Q_R} \,·
}
For the all known theories, $F_{S^{5}}<0$ at large $N$. Then, depending on the sign of $n$ and $Q_{R}$, one chooses the root so that the entropy is positive. Plugging in the solutions for $\lambda$ in any of the equations above leads to a nonlinear constraint among $J_{1,2}$ and $Q_R$.

Taking ratios of the saddle equations leads to the simple equations for the rotational fugacities,
\equ{\label{saddlew1w2}
\frac{\varphi}{\omega_1}= \frac{S-2\pi i n J_1}{S+2\pi i n Q_R} \,,\qquad \frac{\varphi}{\omega_2}= \frac{S-2\pi i n J_2}{S+2\pi i n Q_R} \,·
}
In the next section we discuss holographic applications in the study of black holes.

\section{Black hole entropy}
\label{sec:Black hole entropy}

In this section, we discuss the implications of the large $N$ Cardy  formula in holography, in particular for the microscopic counting of black hole entropy. In the context of holography, the rigid background $S^4\times S^1$ is seen as specifying the boundary of a six-dimensional quantum gravitational theory in asymptotically locally AdS$_6$. In the regime in which the dual bulk theory is weakly coupled, the partition function of the boundary field theory can then be reorganized as a sum over semiclassical gravitational saddles of the bulk theory,
\equ{\label{sumgeom}
Z_{S^4_{\omega_1,\omega_2}\times S^1}(\nu_{I})\approx \sum_{\substack{\text{gravitational} \\ \text{saddles}}} e^{-I_E(\nu_{I};\, \omega_{1},\omega_{2})}\,,
}
where $I_E$ is the Euclidean on-shell action of the gravitational saddle and the  chemical potentials in the field theory are identified with  the asymptotic values of the dual gauge fields defining the boundary conditions. In general, there may be various gravitational saddles contributing to \eqref{sumgeom}, with their relative weight varying over the space of fugacities. An interesting question is  which gravitational saddles precisely contribute and how they can be extracted from the field theory partition function. 

A full understanding of this is out of reach at the moment, requiring both a full classification of the gravitational saddles as well as the corresponding field theory configurations (see discussion at the end of Section~\ref{sec:A large N Cardy formula}). Our claim here is that the contribution leading to \eqref{square} (or \eqref{5dcardyT}) is dual  to the single-center BPS black hole, carrying two angular momenta and multiple electric charges.  For this to be true the contribution \eqref{5dcardyT} to the matrix model should equal the Euclidean on-shell action of the corresponding black hole, i.e.,
\equ{\label{EAuniv}
I_E(\nu_{I};\omega_{1},\omega_{2})= \frac{\varphi^{3}}{i \pi \omega_{1}\omega_{2}}  F_{ S^5}(\nu_{I})\,.
}
The parameter $\varphi$ corresponds to the chemical potential associated to electric charge under the graviphoton and is fixed in terms of the rotational fugacities $\omega_{1,2}$ by supersymmetry. The parameters $\nu_{I}$ should be identified with chemical potentials for additional electric charges. Unfortunately, no spinning black holes with multiple electric charges are known at the moment and we cannot test \eqref{EAuniv} directly. However, in the important case $\nu_{I}=0$ the black hole is known and there is perfect agreement as we discuss in more detail below. 

\

Another quantity of physical interest is, of course, the Bekenstein-Hawking entropy of the Lorentzian solution. This is somewhat subtle since (supersymmetric) Euclidean black holes in AdS do not typically admit a regular continuation to Lorentzian signature unless an additional extremality constraint is imposed, as discussed in    \cite{Cassani:2019mms,Hosseini:2017mds,Cabo-Bizet:2018ehj,Choi:2018fdc,Choi:2019miv} in various dimensions. As in those cases we expect the  superconformal index to be aware of the extremality constraint, as we discuss below.

\subsection{The universal spinning black hole}
\label{sec:The universal spinning black hole}

Let us consider the universal case, $\nu_{I}=0$, discussed at the end of Section~\ref{sec:The superconformal index}. On the supergravity side this corresponds to setting all vector multiplets of the theory to zero, which is a consistent truncation to minimal 6d $F(4)$ gauged supergravity, containing only the gravity multiplet.  A spinning black hole solution in this theory was  constructed in \cite{Chow:2008ip}, carrying two angular momenta and an electric charge under the graviphoton, which is dual to the $SU(2)_{R}$ R-symmetry. The renormalized Euclidean on-shell action of this solution was recently computed in  \cite{Cassani:2019mms}. Reinstating Newton's constant, it is given in terms of the rotational and electric chemical potentials by
\equ{\label{CP}
I_{E}= \frac{i\pi}{3 G_{(6)}} \frac{\varphi^{3}}{\omega_{1}\omega_{2}} =\frac{1}{i\pi} \frac{\varphi^{3}}{\omega_{1}\omega_{2}} F_{S^{5}}\,,
}
where in the second equality we used $F_{S^5} = -\frac{\pi^2}{3 G_{(6)}}$, with $F_{S^5}$  the free energy of the {\it undeformed} SCFT on the round $S^{5}$, and $3\varphi=\nu_{R}$ is identified with the  R-symmetry fugacity in the superconformal index \eqref{SCIdeg}. This precisely matches the field theory result \eqref{EAuniv} for $\nu_{I}=0$.  Furthermore,  this match holds for finite $\omega_{1,2}$ and irrespective of the uplift of the 6d solution, either to massive type IIA or type IIB string theory,  as argued in \cite{Bobev:2019zmz}.\footnote{The information of the particular uplift is encoded in the expression for Newton's constant $G_{(6)}$ in terms of the 10d geometry, which can be obtained  by the uplift results of \cite{Cvetic:1999un} for massive type IIA and \cite{Jeong:2013jfc,Hong:2018amk,Malek:2018zcz,Malek:2019ucd} for type IIB.} This justifies our claim that the eigenvalue configuration \eqref{saddleS5}, leading to \eqref{5dcardyT} is dual to a single-center black hole. 

We note that the eigenvalue configuration \eqref{saddleS5} exists for generic $\omega_{1,2}$, in particular, in the Cardy-like limit $|\omega_{1,2}|\ll1$ studied in  \cite{Choi:2019miv}, which corresponds to black holes that are large compared to the size of AdS. Thus, we have shown the saddle persists beyond the Cardy-like limit as the black hole becomes smaller, thus accounting for the extended validity of the 5d Cardy formula. It is possible, however,  that when the black hole shrinks beyond a critical size, other supergravity configurations with the same asymptotic charges may exist and even dominate the canonical ensemble, leading to a corresponding phase transition in the field theory. It would be interesting to study if such supergravity configurations exist. From now on we focus on the single-center black hole. 

The  match between the field theory free energy and the supergravity on-shell action of the single-center black hole is sufficient to give a microscopic derivation of the black hole's entropy, and regardless of its size. This is because the  supersymmetric black hole satisfies the quantum statistical relation \cite{Chow:2008ip} (see \cite{Choi:2018fdc,Cassani:2019mms} for a detailed discussion of the supersymmetric case),\footnote{Our definition of $Q$ differs by a factor 3 compared to these references.}
\equ{\label{BHentropy}
S_{\text{BH}}= -I_{E}-\omega_{1}J_{1}-\omega_{2}J_{2}-3\varphi Q\,,
}
where $S_{\text{BH}}=\frac{\text{Area}}{4G_{(6)}}$ is the Bekenstein-Hawking entropy. Setting $\nu_{I}=0$ in \eqref{FTentropyE} and comparing to \eqref{BHentropy} it is then {\it automatic} that the field theory degeneracy reproduces the Bekenstein-Hawking entropy:  $S_{\text{BH}}=\log \Omega$.\footnote{As discussed in Section~\ref{sec:microcanonicalpf}, here we assume the constraint among $J_{1,2}$ and $Q_{R}$ that ensures the entropy is real.} The fact that the black hole entropy can be obtained by the Legendre transform \eqref{BHentropy} was first observed in \cite{Cassani:2019mms,Choi:2018fdc}. The field theoretical derivation above is an explanation of this observation.

We emphasize that $S_{\text{BH}}=\log \Omega$ holds regardless of the uplift to 10d. This simple observation accounts for the microscopic entropy of an {\it infinite} class of black holes, large or small, in both massive type IIA and type IIB string theory.

\subsection{A mesonic spinning black hole in massive type IIA}

We now consider an example with nonzero flavor fugacities. As a simple example  we consider the  Seiberg theory  \cite{Seiberg:1996bd}, consisting of a  single 5d $\cN=1$ vector multiplet in the adjoint of the gauge group $Sp(N)\simeq USp(2N)$, $N_{f}\leq 7$  hypermultiplets in the fundamental representation, and one hypermultiplet in the antisymmetric representation.  The global flavor symmetry is $SU(2)_{M}\times SO(2N_{f})\times U(1)_{\mathit{top}}$, where the first factor acts on the antisymmetric hypermultiplet, the second factor on the fundamental  hypermultiplets, and the last factor is a topological symmetry. In full generality, one may turn on all fugacities in the Cartan of this large flavor group.  For simplicity here we turn on a fugacity only for the Cartan of the mesonic $SU(2)_M$, which is translated by the map \eqref{5dcardyT} into turning on a real mass for this symmetry in the $S^5$ partition function. The mass-deformed partition function on the round $S^5$  was computed in \cite{Chang:2017mxc} (see also \cite{Gutperle:2018axv,Hosseini:2019and}):
\equ{F_{S^5}(\hat{m}_{\text{AS}})=- \frac{\sqrt 2 \pi(9+4\hat{m}_{\text{AS}}^2)^{3/2}}{15 \sqrt{8-N_f}} N^{5/2}+\mathcal O(N^{3/2})\,,
}
where $\hat{m}_{\text{AS}}$ is the mass parameter for the antisymmetric hypermultiplet. With $\nu_{AS}=\varphi\,\hat{m}_{AS}$ and 
\equ{\Delta_1 = 3\varphi +2 i \nu_{AS}\,,\qquad  \Delta_2 = 3\varphi -2 i \nu_{AS}\,,
}
we have
\equ{\frac{\varphi^{3}}{i \pi \omega_{1}\omega_{2}} F_{S^5}(\hat{m}_{\text{AS}})= \frac{\(\Delta_1\Delta_2\)^{3/2}}{27 i \pi \omega_{1}\omega_{2}} F_{S^5}\,,
}
where $F_{S^5}\equiv F_{S^5}(\hat{m}_{\text{AS}}=0)$ is the free energy at the 5d conformal point.

The field theoretical entropy is now determined by extremizing
\equ{S= \frac{\(\Delta_1\Delta_2\)^{3/2}}{27 i \pi \omega_{1}\omega_{2}} F_{S^5} - \omega_1 J_1-\omega_2 J_2-\Delta_1 Q_1-\Delta_2 Q_2  +\lambda(\omega_1+\omega_2-\tfrac{1}{2}(\Delta_1+\Delta_2) +2\pi  i n)\,,
}
with respect to the chemical potentials $\omega_{1,2}$ and  $\Delta_{1,2}$ and  the Lagrange multiplier $\lambda$, which  we have included to ensure the supersymmetric constraint on the index. This is consistent with the proposal in \cite{Zaffaroni:2019dhb} and the derivation in the large $N$ and $|\omega_{1,2}|\ll 1$ limit in \cite{Choi:2019miv}. This leads to the equations
\eqsn{0=\,&\lambda-J_1 - \frac{\(\Delta_1\Delta_2\)^{3/2}}{27 i \pi \omega_{1}^2\omega_{2}} F_{S^5} \,,&& 0= \lambda-J_2 - \frac{\(\Delta_1\Delta_2\)^{3/2}}{27 i \pi \omega_{1}\omega_{2}^2} F_{S^5} \,,\\
0=\,& \frac12 \lambda+Q_1 -\frac{3}{2\Delta_1} \frac{\(\Delta_1\Delta_2\)^{3/2}}{27 i \pi \omega_{1}\omega_{2}} F_{S^5} \,,&& 0=  \frac12 \lambda+Q_2 -\frac{3}{2\Delta_2} \frac{\(\Delta_1\Delta_2\)^{3/2}}{27 i \pi \omega_{1}\omega_{2}} F_{S^5}\,,
}
and the constraint $\omega_1+\omega_2-\frac12(\Delta_1+\Delta_2) +2\pi  i n=0$. At the extremum,
\equ{\label{slambda}
S= 2 \pi i n \lambda\,.
}
Thus, to determine the entropy it suffices to find $\lambda$. Taking products of the equations above one can easily write an equation for $\lambda$ in terms of the charges:
\equ{\label{eql}
(\lambda-J_1)^{2}(\lambda-J_2)^{2}F_{S^5}^{2} = -64 \pi^{2} \(\frac{\lambda}{2}+Q_1\)^{3}\(\frac{\lambda}{2}+Q_2\)^{3}\,.
}
The solution to this sixth-order polynomial equation for $\lambda$  is generically complex and thus the entropy \eqref{slambda} is not real. Requiring  a purely imaginary value for $\lambda$ and taking the complex conjugate equation (assuming $Q_{1,2}$ and $J_{1,2}$ real), we get
\equ{\label{eqlc}
(\lambda+J_1)^{2}(\lambda+J_2)^{2}F_{S^5}^{2} = -64 \pi^{2} \(\frac{\lambda}{2}-Q_1\)^{3}\(\frac{\lambda}{2}-Q_2\)^{3}\,.
} 
Taking the difference of \eqref{eql} and \eqref{eqlc}, and assuming $\lambda \neq 0$, we have
\eqst{3 c  (Q_1+Q_2)\lambda^4 -4\left[16 (J_1+J_2) -c(Q_1+Q_2)(Q_1^2+8 Q_1Q_2+Q_2^2)\right]\lambda^2 \\
-16 \left[4 J_1 J_2 (J_1+J_2)-3cQ_1^2Q_2^2(Q_1+Q_2))\right]=0\,,
}
where $c\equiv \frac{64 \pi^{2}}{F_{S^5}^2}\,·$ This equation  can be easily solved for $\lambda$, which determines the entropy via \eqref{slambda}.  Then, plugging the solution into \eqref{eql} or \eqref{eqlc} leads to a nonlinear constraint among the angular momenta and electric charges that ensures the entropy is real. Based on similar analyses for other spinning black holes in AdS$_{5}$ \cite{Hosseini:2017mds,Cabo-Bizet:2018ehj} and AdS$_{4}$ \cite{Choi:2018fdc,Cassani:2019mms} we expect this constraint to correspond to an extremal (zero temperature) limit.\footnote{See \cite{Zaffaroni:2019dhb} for a pedagogical exposition and review of existing results.} It would be interesting to construct this solution, derive the extremal constraint, and match the entropy above.

\section{\texorpdfstring{The $\bm{S^3_{b}\times \Sigma_{\fg}}$ partition function}{The S³b×Σg Partition Function}}
\label{sec:S3Sigma}

In this section, we briefly review the results \cite{Crichigno:2018adf} for the $S^{3}_b\times \Sigma_{\fg}$ partition function and observe that the results there can be written compactly in terms of the squashed $S^{5}$ partition function. For a general theory with gauge group $G$ and hypermultiplets in gauge representations $\cR_{I}$ the perturbative partition function is given by:
\eqss{\label{ZS3bpert}
Z^{\mathit{pert}}_{S^{3}_{b}\times \Sigma_{\fg}} = \sum_{ \hat u \in \mathcal S_{\mathit{BE}}}\,  H^{\fg-1} &\prod_{\alpha\in \mathit{Ad}(G)'} s_{b}\(\alpha(\hat{u})-iQ\)^{1-\fg} \,\prod_{I}\prod_{\rho\in \cR_{I}}s_{b}\(\rho(\hat{u})+\nu_{I}\)^{(\fg-1) \hat \fn_{I}}\,,
}
where $s_b(x)$ is the double sine function, $Q=\frac12(b+b^{-1})$, $\nu_I$ and $\hat \fn_I$ are a fugacity and a quantized flux for flavor symmetry, respectively,  and the sum is over solutions to the Bethe equations  
\equ{\label{SBESec3}
\cS_{\mathit{BE}} = \;\; \left \{ \hat{u} \;\; \big| \;\; \Pi_a(\hat{u}) \equiv \exp \( 2 \pi i \frac{\partial \cW^{\mathit{pert}}_{S^3_b \times \R^2}}{\partial u_a}(\hat{u})\) = 1, \;\;\; a=1,...,r_G\right\} /W_G\,,
}
where $\mathcal W^{\mathit{pert}}_{S^{3}_{b}\times \mathbb R^{2}}$ is the perturbative twisted superpotential, given by
\equ{\label{Bethepert}
\mathcal W^{\mathit{pert}}_{S^{3}_{b}\times \mathbb R^{2}}(\tu,\tnu) =\mathcal W^{\mathit{classical}}_{S^{3}_{b}\times \mathbb R^{2}}(\tu,\tnu) +\sum_{I}\sum_{\rho\in \cR_{I}}g_b\big(\rho(\tu)+\tnu_{I}\big)-\sum_{\alpha∈Ad(G)'}g_b\big( \alpha(\tu)+1\big) \,.
}
The variables with and without ``tilde'' are related by $\tu= i Q^{-1}u$ and $\tnu=i Q^{-1}\nu$.  The classical contribution is given by
\equ{\label{wclass}
\cW^{classical}_{S^3_b \times \R^2}(\tu) =- \frac{1}{2} Q^2 \tgamma \tu^2 - \frac{1}{6} Q^2\text{Tr}_{\mathit CS} (\tu^3)\,,
}
and the functions $s_{b}(x)$ and $g_b(x)$ are related by
\equ{\log s_b(-iQx)=2πi g'_b(x)\,.
}
To study the large $N$ limit, we will need only the asymptotic behavior of $g_b(x)$, as $\tu→± i ∞$,
\equ{g_{b}(\tu+\tnu) \rightarrow \pm\Bigg(\!\!-\frac{1}{12}Q^{2}\tu^{3}+\frac14Q^{2}\tnu\tu^{2}+\frac14\left[\frac{1+4Q^{2}}{6}-Q^{2}(1-\tnu^{2})\right]\tu\Bigg)\,.
\label{expansiongb}}
The function $H$ is a Hessian contribution, which is subleading in the large $N$ limit and hence we ignore (see \cite{Crichigno:2018adf} for details). Similarly, we assume instanton corrections are suppressed at large $N$ in the remainder of this section we shall always omit the label ``{pert}''  to avoid clutter.

The large $N$ limit of this observable was studied in \cite{Crichigno:2018adf}. Here we simply point out that the results there can be written compactly as\footnote{Similar results appear for topologically twisted indices in 3d \cite{Hosseini:2016tor,Jain:2019lqb} and in 5d \cite{Hosseini:2018uzp}.}
\eqs{\mathcal W_{S^{3}_{b}\times \mathbb R^{2}}(\tnu) \,&\approx \frac{4}{27\pi}F_{S^5_{\vec{ω}=(b,b^{-1},Q)}}\(\frac{3Q\tnu}{2}\), \\
\log Z_{S^{3}_{b}\times \Sigma_{\fg}}(\tnu)_{\hat \fn} \,&\approx 2\pi (\fg-1)\(3\mathcal W_{S^{3}_{b}\times \mathbb R^{2}}(\tnu)+\sum_I \(\hat \fn_I-\tnu_I\)\frac{\partial \mathcal W_{S^{3}_{b}\times \mathbb R^{2}}(\tnu)}{\partial \tnu_I}\),
\label{S3Sigmarels}}
where $F_{S^5_{\vec{ω}}}\(m\)$ is the free energy of the theory on $S^5_{\vec{ω}}$, deformed by a mass parameter $m$. Here $\mathcal W_{S^{3}_{b}\times \mathbb R^{2}}(\tnu)\equiv \mathcal W_{S^{3}_{b}\times \mathbb R^{2}}(\hat \tu,\tnu)$ denotes the twisted superpotential, evaluated at the  vacua \eqref{SBESec3}. The universal twist corresponds to setting
\equ{\hat \fn_I=0 = \tnu_I\,,
}
which in particular implies 
\equ{\label{universal89}
\log Z_{S^{3}_{b}\times \Sigma_{\fg}}\approx -\frac{8}{9}(\fg-1)Q^2 \log Z_{S^{5}}\,.
}
This universal relation between partition functions holds for any 5d theory with a weakly coupled gravity dual, including theories engineered in both massive type IIA and type IIB string theory. 

\paragraph{Proof of (\ref{S3Sigmarels}).} We begin by rewriting \eqref{ZS3bpert} as (ignoring the Hessian contribution):
\eqst{\log Z_{S^{3}_{b}\times \Sigma_{\fg}} = -∑_{\alpha\in \mathit{Ad}(G)'} 2πi(\fg-1)g'_b(α(\hat{\tu})+1) +∑_{I}∑_{\rho\in  R_{I}}2πi(\fg-1)\hat\fn_{I}g'_b(ρ(\hat{\tu})+\tnu_I) \\
=∑_{I}∑_{\rho\in  R_{I}}2πi(\fg-1)\big(\hat\fn_{I}-\tnu_I^{-1}\big)g'_b(ρ(\hat{\tu})+\tnu_I) =2π(\fg-1)∑_I\big(\hat\fn_{I}-\tnu_I^{-1}\big)\frac{∂\W_{S^3_b×\R^2}(\tnu)}{∂\tnu_I} \,,
\label{proof1}}
where the second line follows by using the large $N$ constraint \eqref{constquiver} to solve the Bethe equations \eqref{SBESec3} (with the classical contribution in \eqref{Bethepert} being considered subleading) and the overall $i$ is removed as the prime stands for $\frac{\partial}{\partial(i\tu)} =-i\frac{∂}{\partial{\tnu}}\,·$ It is useful to introduce an extended set of fugacities, $\nu_{I}^{\pm}$, defined by\footnote{This definition is so that the coefficient $1-\tnu^{2}=\tnu^{+}\tnu^{-}$ in \eqref{expansiongb} is homogeneous in these variables.}
\equ{\tnu_{I}^{\pm}\equiv 1\pm \tnu_{I}\,,\qquad \tnu_{I}^{+}+\tnu_{I}^{-}=2\,.
}
The sum appearing in \eqref{proof1} runs over an independent set of $\tnu$'s. We can rewrite the $\hat{\fn}_I$-independent piece in terms of the extended set by using the following summation relation\footnote{This relation follows by differentiating $\W(\tnu)$ and imposing the constraint on $\tnu$ in different order. A similar trick is used to prove the relation for 3d topologically twisted index in \cite{Jain:2019lqb}.}
\equ{\sum_{I}∑_{\pm}\tnu_{I}^{\pm}\frac{∂\W_{S^3_b×\R^2}(\tnu)}{∂\tnu_{I}^{\pm}} =∑_I\(\tnu_I -\tnu_I^{-1}\)\frac{∂\W_{S^3_b×\R^2}(\tnu)}{∂\tnu_I}\,·
\label{proof2}}
In addition, one can see that $\W_{S^3_b×\R^2}(\tnu)$ is homogeneous of degree 3 in the extended variables $\tnu_{I}^{\pm}$, i.e., 
\equ{\sum_{I}∑_{\pm}\tnu_{I}^{\pm}\frac{∂\W_{S^3_b×\R^2}(\tnu)}{∂\tnu_{I}^{\pm}} =3\W_{S^3_b×\R^2}(\tnu)\,.
\label{proof3}}
To show this, focus on \eqref{Bethepert} (ignoring the subleading classical contribution) where the functions $g_{b}$ are expanded using \eqref{expansiongb}. The important points are that the quadratic term in $\tu$ cancels out to leading order and, assuming \eqref{constquiver}, the linear terms in $\tu$ from the vector and hypermultiplet combine into the form $ \(\sum_{I}\tnu_{I}^{+}\tnu_I^{-}\)\tu$. Thus, only a cubic and this linear term remain. Then, rescaling $\tu\to \(\sum_{I}\tnu_{I}^{+}\tnu_I^{-}\){\!}^{1/2}\tu$ brings the dependence on the fugacities into an overall factor  of $\(\sum_{I}\tnu_I^{+}\tnu_I^{-}\){\!}^{3/2}$ in the twisted superpotential, which proves the claim. This holds for theories in both massive type IIA and type IIB string theory.\footnote{See \cite{Crichigno:2018adf} for explicit examples in massive type IIA.}  Finally, combining \eqref{proof1}, \eqref{proof2} and \eqref{proof3}, we obtain \eqref{S3Sigmarels}.

\paragraph{Comment on novel 3d SCFTs and holography.} The $S^3_b\times \Sigma_{\fg}$ partition function can be interpreted as the $S^{3}_{b}$ partition function of the 3d SCFTs obtained by twisted compactification of the 5d SCFTs on $\Sigma_\fg$ and flowing to the IR. These 3d theories are then labelled by the UV theory one started with in 5d and the compactification data such as discrete fluxes, $\fn$, on the Riemann surface:
\eqss{\cT^{(5d)}	\rightsquigarrow  \cT^{(3d)}_{\Sigma_{\fg},\fn}	\,,\qquad
Z_{S^{3}_{b}\times\Sigma_{\fg}}\Big[\cT^{(5d)}\Big]_{\fn} =  Z_{S^{3}_{b}}\Big[\cT^{(3d)}_{\Sigma_{\fg},\fn}\Big]\,.
}
Determining whether this procedure actually leads to interacting 3d SCFTs in the IR is a nontrivial problem. At large $N$ one can gain some insights into this question as the explicit RG flow from 5d to 3d can be constructed holographically as a solution of 6d $F(4)$ minimal gauged supergravity interpolating between locally AdS$_6$ and AdS$_4\times \Sigma_\fg$. As shown  in \cite{Bobev:2017uzs} properties of the supergravity solution imply the relation $F_{S^{3}\times \Sigma_{\fg}}= -\frac{8}{9}(\fg-1)  F_{S^{5}}$, which  exactly matches the large $N$ relation  \eqref{universal89}, for the round sphere $b=1$.  This strongly suggests that the fixed points $\cT^{(3d)}_{\Sigma_{\fg},\fn}$  are nontrivial and strongly interacting, both in the massive type IIA and type IIB constructions. Note the scaling with $N$ of the 3d partition function is inherited from 5d and thus is given by  $N^{5/2}$ and $N^{4}$, respectively.  It would be interesting to uncover whether these theories admit a purely three-dimensional description.

\subsection*{Acknowledgements}

We would like to thank Nikolay Bobev,  Zohar Komargodski, and  Guli Lockhart for discussions. The work of PMC is supported by Nederlandse Organisatie voor Wetenschappelijk Onderzoek (NWO) via a Vidi grant and is also part of the Delta ITP consortium, a program of the NWO that is funded by the Dutch Ministry of Education, Culture and Science (OCW).

\appendix

\section{Superconformal algebra}
\label{appAlgebra}

The superconformal algebra contains the supercharges $\cQ^A_m$ and $\cS_A^m$, which in radial quantization are conjugates of each other,  $\cS_A^m=(\cQ^A_m)^\dagger$.  They satisfy \cite{Bhattacharya:2008zy} (here we follow conventions in \cite{Kim:2012gu})
\equ{\{\cQ_m^A,\cS_B^n\} = \delta_m^n\delta_B^A D +2 \delta_B^AM_m^{\;\;n}-3 \delta_m^nR_B^{\;\;A}\,,
}
where $D$ is the generator of dilatations, $M_m^{\;\;n}$ are the generators of $SO(5)$ rotations, and $R_B^{\;\;A}$ are generators of  $SU(2)_R$. Explicitly, 
\eqss{\{\cQ^1_1,(\cQ^1_1)^\dagger\}=\,&\Delta +J_{1}+J_{2}-3R\,,\\
\{\cQ^1_2,(\cQ^1_2)^\dagger\}=\,&\Delta -J_{1}-J_{2}-3R\,,\\
\{\cQ^2_2,(\cQ^2_2)^\dagger\}=\,&\Delta -J_{1}+J_{2}+3R\,,\\
\{\cQ^2_1,(\cQ^2_1)^\dagger\}=\,&\Delta +J_{1}-J_{2}+3R\,.
}
One now chooses a supercharge, $\cQ$, out of the set above and defines the corresponding index which is, by construction, invariant  under the supercharge $\cQ$. The index defined in \eqref{Hindex} corresponds to choosing $\cQ \equiv Q_{2}^{1}$ and  thus receives contributions only from states satisfying
\equ{\Delta -J_{1}-J_{2}-3R=0\,.
}
Imposing this relation in the other three anticommutators, and assuming unitarity, implies that the states contributing to the index also satisfy
\eqss{J_1+3 R\geq 0\,, \qquad J_2+3 R\geq 0\,, \qquad J_1+J_2\geq 0\,.
}
If any of these inequalities is saturated the state preserves additional supersymmetry.

\section{Special functions and asymptotics}
\label{App:FunIds}

Here we collect some results for the special functions used in the main text, mostly following \cite{FELDER200044}, and derive the asymptotic behavior of the elliptic gamma function. 

We begin with the $\fq$-Pochhammer symbol $(z;\fq)$,  defined by
\equ{(z;\fq)\equiv \begin{cases}\prod_{k\geq 0}(1-z\fq^{k})\qquad &|\fq|<1 \\
\prod_{k\geq 0}(1-z\fq^{-k-1})^{-1}\qquad &|\fq|>1
\end{cases}\,,
}
where $z$ and $\fq$ are complex variables. It satisfies the following properties, for $|\fq|≠1$,
\equ{(\fq z;\fq) =(1-z)^{-1}(z;\fq)\,, \qquad (z;\fq^{-1}) =(\fq z;\fq)^{-1}\,.
}
The $\fq$-Pochhammer  symbol can be used to define the elliptic theta function:
\equ{θ(z;\fq)\equiv (z;\fq)(\fq z^{-1};\fq)\,.
}

One can define multiple $\vec \fq$-Pochhammer  symbols. In particular, the double $(\fq_{1},\fq_{2})$-Pochhammer  symbol is defined as
\equ{(z;\fq_{1},\fq_{2})\equiv \begin{cases}\prod_{k_{1},k_{2}\geq0}(1-z\fq_{1}^{k_{1}}\fq_{2}^{k_{2}})\qquad &|\fq_{1}|<1\, ,|\fq_{2}|<1 \\
\prod_{k_{1},k_{2}\geq0}(1-z\fq_{1}^{-k_{1}-1}\fq_{2}^{-k_{2}-1})\qquad &|\fq_{1}|>1\, ,|\fq_{2}|>1 \\
\prod_{k_{1},k_{2}\geq0}(1-z\fq_{1}^{k_{1}}\fq_{2}^{-k_{2}-1})^{-1}\qquad &|\fq_{1}|<1\, ,|\fq_{2}|>1 \\
\prod_{k_{1},k_{2}\geq0}(1-z\fq_{1}^{-k_{1}-1}\fq_{2}^{k_{2}})^{-1}\qquad &|\fq_{1}|>1\, ,|\fq_{2}|<1
\end{cases}\,.
}
With these definitions, various identities follow:
\eqsg{(z;\fq_1,\fq_2)=(z\fq_1^{-1};\fq_1^{-1},\fq_2)^{-1}\,,&\quad (z;\fq_1,\fq_2)=(z\fq_1^{-1}\fq_2^{-1};\fq_1^{-1},\fq_2^{-1})\,, \\
(\fq_1 z;\fq_1,\fq_2)=(z;\fq_2)^{-1}(z;\fq_1,\fq_2)\,,&\quad (\fq_2 z;\fq_1,\fq_2)=(z;\fq_1)^{-1}(z;\fq_1,\fq_2)\,.
\label{idpocha}}
The elliptic gamma function is defined as
\equ{\Gamma\(z;\fq_{1},\fq_{2}\)=\frac{(\fq_{1}\fq_{2}z^{-1};\fq_{1},\fq_{2})}{(z;\fq_{1},\fq_{2})}\,·
\label{DefEllipticGammaA}}
We often use the shorthand notation
\equ{\Gamma(u;\ep_{1},\ep_{2})\equiv \Gamma\big(z=e^{2\pi i u};\fq_{1}=e^{2\pi iε_1},\fq_{2}=e^{2\pi iε_2}\big)\,.}
It satisfies the following ``shift'' properties:
\eqs{Γ(u+1;ε_1,ε_2) =\,&Γ(u;ε_1,ε_2) \,,\\
Γ(u+ε_1;ε_1,ε_2) =\,& θ(u;ε_2)Γ(u;ε_1,ε_2)\,,\\
 Γ(u+ε_2;ε_1,ε_2) =\,& θ(u;ε_1)Γ(u;ε_1,ε_2)\,, \\
Γ(ε_i -u;ε_1,ε_2) =\,&θ(u;ε_i)^{-1}Γ(u;ε_1,ε_2)^{-1}\,, \\
Γ(ε_1+ε_2-u;ε_1,ε_2) =\,& Γ(u,ε_1,ε_2)^{-1}\,,
}
and the ``inversion'' properties:
\equ{Γ(u;-ε_1,ε_2) =Γ(ε_2-u;ε_1,ε_2)\,,\qquad Γ(u;ε_1,-ε_2) =Γ(ε_1-u;ε_1,ε_2)\,.
}
All these formulae hold for  $u∈\bC$ and $ε_i∈\bC\backslash \bR$. The elliptic gamma function also satisfies a ``modular'' property for $ε_i,\frac{ε_2}{ε_1},ε_1+ε_2 ∈\bC\backslash \bR$ \cite{FELDER200044}:
\equ{Γ(u;ε_{1},ε_{2})=e^{-i π Q(u;ε_1,ε_2)}\frac{Γ\big(\frac{u}{ε_1};-\frac{1}{ε_1},\frac{ε_2}{ε_1}\big)}{Γ\big(\frac{u-ε_1}{ε_2};-\frac{1}{ε_2},-\frac{ε_1}{ε_2}\big)}\,,
\label{GammaIdA}}
where  
\eqs{Q=\frac{u^3}{3 ε_1 ε_2}-\frac{ε_1+ε_2-1}{2 ε_1 ε_2}u^2 +\frac{(ε_1+ε_2-1)^2+ε_1 ε_2-ε_1-ε_2}{6 ε_1 ε_2}u
-\frac{(ε_1+ε_2-1) (ε_1ε_2-ε_{1}-ε_{2})}{12 ε_1 ε_2}\,·
\label{defQA}}

Another relevant function is  Barnes' multiple gamma function, defined as
\equ{Γ_M(u|\vec{ω}) =\exp\(\left.\frac{∂ζ_M(s,u|\vec{ω})}{∂s}\right|_{s=0}\);\qquad ζ_M(s,u|\vec{ω}) =∑_{\vec{n}=0}^∞\frac{1}{(u+n_1ω_1+\cdots+n_Mω_M)^s}\,·
}
The asymptotics of  Barnes' multiple gamma function as $|u|\to \infty$ was studied by Ruijsenaars in \cite{RUIJSENAARS2000107}. It was shown there that $Γ_M(u|\vec{ω})$ admits the expansion
\equ{\label{asymptGM}
\log Γ_M(u|\vec{ω}) = \frac{(-1)^{M+1}}{M!}B_{M,M}(u)\log u +(-1)^{M}\sum_{n=0}^{M-1}\frac{B_{M,n}(0)u^{M-n}}{n! (M-n)!}\sum_{l=1}^{M-n}\frac{1}{l}+\mathcal O\(\frac{1}{|u|}\),
}
which holds for   $\Re\, u>0$,  $\Re \, ω_i>0$,  $|\arg(u)|<π$,  and  where  $B_{M,n}(u|\vec{ω})$ are the multiple Bernoulli polynomials, defined by the generating function
\equ{\frac{t^Me^{ut}}{∏_{i=1}^M\(e^{ω_i t} -1\)} =∑_{n=0}^{∞}\frac{t^n}{n!}B_{M,n}(u|\vec{ω})\,.
}
The $B_{M,n}(u|\vec{ω})$ are polynomials of degree $n$ in the variable $u$ and symmetric in $ω_1,⋯,ω_M$. Two useful properties are the rescaling and shift properties, \cite{NARUKAWA2004247}
\equ{B_{M,n}(r u|r\vec{ω})=r^{n-M}B_{M,n}(u|\vec{ω})\,,\qquad B_{M,n}(ω_{\text{tot}}-u|\vec{ω}) =(-1)^nB_{M,n}(u|\vec{ω})\,,
}
where $ω_{\text{tot}}=∑_{i=1}^Mω_i$. The relevant polynomials for our purposes are, explicitly:
\eqs{B_{3,0}\(u|\vec{\omega}\) &=\frac{1}{ω_1ω_2ω_3}\,,\qquad B_{3,1}\(u|\vec{\omega}\) =\frac{u}{ω_1ω_2ω_3} -\frac{ω_{\text{tot}}}{2ω_1ω_2ω_3}\,, \nn
B_{3,2}\(u|\vec{\omega}\) &=\frac{u^2}{\omega_1\omega_2\omega_3} -\frac{\omega_{\text{tot}}}{\omega_1\omega_2\omega_3}u +\frac{\omega_{\text{tot}}^2 +\omega_1\omega_2+\omega_1\omega_3+\omega_2\omega_3}{6\omega_1\omega_2\omega_3}\,, \label{defB}\\
B_{3,3}\(u|\vec{\omega}\) &=\frac{u^3}{\omega_1\omega_2\omega_3} -\frac{3\omega_{\text{tot}}}{2 \omega_1\omega_2\omega_3}u^2 +\frac{\omega_{\text{tot}}^2 +\omega_1\omega_2+\omega_1\omega_3+\omega_2\omega_3 }{2\omega_1\omega_2\omega_3}u - \frac{\omega_{\text{tot}}(\omega_1\omega_2+\omega_1\omega_3+\omega_2\omega_3)}{4\omega_1\omega_2\omega_3}\,·\nonumber
}

\subsection{Asymptotics of the elliptic gamma function}
\label{sec:AsymptoticsEG}

Combining the results reviewed above one can easily derive the asymptotics of the elliptic gamma function, which controls the large $N$ limit of the superconformal index.  The crucial relation for us is an identity relating Barnes' triple gamma function, $Γ_3(u|\vec{ω})$, to the elliptic gamma function, $Γ(u;ε_1,ε_2)$. Assuming $\Im\, ε_i>0$ one can show that (see Corollary 6.2 in \cite{FRIEDMAN2004362}):
\equ{\label{triplegammaA}
Γ(u;ε_1,ε_2)=e^{iπ R(u-\ep_+|ε_1,ε_2)}\frac{Γ_3(u|\,ε_1,ε_2,1)Γ_3(1-u|-ε_1,-ε_2,1)}{Γ_3(-u+2\ep_+|\,ε_1,ε_2,1)Γ_3(1+u-2\ep_+|-ε_1,-ε_2,1)}\,,
}
where $\ep_+=\half(ε_1+ε_2)$ and  $R$ is the cubic polynomial 
\equ{\label{defRA}
R(v|ε_1,ε_2)=-\frac{1}{6}\(B_{3,3}(v+\ep_+|\,ε_1,ε_2,1)-B_{3,3}(-v+\ep_+|\,ε_1,ε_2,1)\) =-\frac{v^3}{3ε_1ε_2} +\frac{ε_1^2+ε_2^2 -2}{12ε_1ε_2}v\,.
}
Explicitly, using \eqref{asymptGM} for $M=3$ and \eqref{defB}, one has
\eqss{\label{expPsi3}
\log \Gamma_3(u|\vec{ω})\approx\frac{1}{3!}B_{3,3}(u|\vec{ω})\log u -\frac{11}{36 ω_1ω_2ω_3}\,u^3 +\frac{3 ω_{\text{tot}}}{8ω_1ω_2ω_3}\, u^2 -\frac{ω_{\text{tot}}^2+ω_1ω_2+ω_1ω_3+ω_2ω_3}{12 ω_1ω_2ω_3}\,u \,.
}
Using this asymptotic expression for each of the factors in \eqref{triplegammaA}, and choosing the  branch $\log(-u)=-iπ+\log (u)$, one sees that the contributions from first term in \eqref{expPsi3} combine with the polynomial $R$ into a single Bernoulli polynomial, $B_{3,3}(-u|\,ε_1,ε_2,r^{-1})$. On the other hand, the remaining polynomial contributions in \eqref{expPsi3}  mostly cancel out, leaving only a linear term in $u$. Precisely, one finds that in the limit $|u|\to \infty$,
\equ{Γ(x\, \fq_{1}\fq_2;\fq_{1},\fq_{2}) ≡Γ(ru+rε_1+rε_2;rε_1,rε_2) 
\approx e^{\frac{iπ}{3}B_{3,3}(-u|\,ε_1,ε_2,r^{-1})+\cL(u)}\,,
}
where $\cL(u)\equiv \frac{2r^{-1}+\ep_{1}+\ep_{2}}{6\ep_{1}\ep_{2}}\, u$. We also used the rescaling property of the Bernoulli polynomial, $B_{3,3}(ru| r \vec\omega)=B_{3,3}(u|\vec \omega)$, and dropped $\mathcal O(1)$ terms. Note that after combining the contribution from the vector and hypermultiplet the linear term $\cL(u)$ is subleading in $N$ for the quivers we consider, due to the requirement \eqref{constquiver}. This is the main asymptotic formula used in the study of the superconformal index at large $N$.

\bibliography{references} 
\bibliographystyle{hephys}

\end{document}